\documentclass[12pt,letterpaper,fleqn]{article}
\usepackage[utf8]{inputenc}
\usepackage{authblk}
\usepackage{geometry}
\usepackage{xcolor}
\usepackage[round]{natbib}
\usepackage{graphicx}
\usepackage{pdflscape}
\usepackage{lscape}
\usepackage{appendix}
\usepackage{setspace}
\usepackage{dashrule}
\usepackage[shortlabels]{enumitem}
\usepackage{longtable}
\usepackage{chngcntr}
\usepackage{afterpage}

\usepackage{colortbl}
\providecommand{\shadeBench}{\rowcolor[RGB]{255, 255, 158}}

\usepackage{threeparttable}
\usepackage{booktabs}

\usepackage{amsmath}
\usepackage{amsfonts}

\usepackage{bm}

\usepackage{enumitem}
\newlist{steps}{enumerate}{1}
\setlist[steps, 1]{label = Step \arabic*:}

\usepackage{dcolumn}
\newcolumntype{d}[1]{D{.}{.}{#1}}


\usepackage{setspace}
\usepackage{caption}
\usepackage{subcaption}
\definecolor{nblue}{HTML}{000660}
\usepackage[colorlinks=true,urlcolor=nblue,linkcolor=nblue,citecolor=nblue]{hyperref}

\makeatletter
\def\thanks#1{\protected@xdef\@thanks{\@thanks
        \protect\footnotetext{#1}}}
\makeatother

\title{\Large \textbf{General Bayesian time-varying parameter VARs for predicting  government bond yields}\normalsize}
\author[1]{{Manfred M. Fischer*\textsuperscript{1}}\thanks{*Corresponding author: Manfred M. Fischer, Vienna University of Economics and Business, \\ Welthandelsplatz 1, A-1020 Vienna, Austria. E-mail: \href{mailto:manfred.fischer@wu.ac.at}{manfred.fischer@wu.ac.at} \\  
\textsuperscript{1} Vienna University of Economics and Business, \textsuperscript{2} University of Salzburg}, Niko Hauzenberger\textsuperscript{1, 2} \vspace{-0.5cm} \\ Florian Huber\textsuperscript{2}, and Michael Pfarrhofer\textsuperscript{2}}
\date{\today}

\begin{document}

\maketitle\thispagestyle{empty}\normalsize\vspace*{-2cm}\small

\vspace{1.5cm}
\begin{center}
\begin{minipage}{0.8\textwidth}
\noindent \small Time-varying parameter (TVP) regressions  commonly assume that time-variation in the coefficients is determined by a simple stochastic process such as a random walk. While such models are capable of capturing a wide range of dynamic patterns, the true nature of time variation might stem from other sources, or arise from different laws of motion. In this paper, we propose a flexible  TVP VAR  that assumes the TVPs to depend on a panel of partially latent covariates. The latent part of these covariates differ in their state dynamics and thus capture smoothly evolving or abruptly changing coefficients. To determine which of these covariates are important, and thus to decide on the appropriate state evolution, we introduce Bayesian shrinkage priors to perform model selection. As an empirical application, we forecast the US term structure of interest rates and show that our approach performs well relative to a set of competing models. We then show how the model can be used to explain  structural breaks in coefficients related to the US yield curve.
\end{minipage}
\end{center}

\normalsize\renewcommand{\thepage}{\arabic{page}}
\doublespacing
\newpage
\section{Introduction}
Time-varying parameter vector autoregressive (TVP-VAR) models are commonly used in finance and macroeconomics to capture dynamic relations across variables, regime shifts and/or structural changes in economic processes \citep[see][]{primiceri2005,cogley2005drifts,dh2012}. These models typically assume that the parameters evolve over time according to a simple stochastic process such as a random walk.  While being rather flexible and parsimonious, this assumption does not allow to examine the extent to which covariates cause changes in the time-varying parameters (TVPs) over time. Moreover, wrongly assuming a random walk state equation could negatively impact predictive accuracy because it essentially implies a smoothness prior on the coefficients. This might be at odds with the rapid shifts we have observed in financial time series such as bond yields and thus could negatively impact predictive accuracy.

The literature has dealt with the issue of selecting the appropriate law of motion for time-varying parameters (TVPs) by estimating different models separately and then using model selection criteria to discriminate between competing specifications \citep[see, e.g.,][]{sims2006were, koop2009evolution, h2020}. But, to the best of our knowledge, no attempt has been made to develop models that rely on a large set of competing laws of motion for the coefficients and decide which one describes the data best.

This paper proposes a flexible approach to TVP-VARs that efficiently integrates out uncertainty with respect to the state evolution equation. The  approach assumes that the TVPs depend on a potentially large panel of covariates. These covariates are commonly labeled effect modifiers which can be partially latent and might feature their own state equations. In case they are observed, we obtain a model that is closely related to the varying coefficient model originally proposed in \cite{hastie1993varying}. The main advantage of using observed, as opposed to latent effect modifiers, is that we can investigate the driving sources of parameter change. This feature is important if the researcher is interested in investigating why relations between variables in a VAR change over time, and to what extent these changes are explained by the observed effect modifiers.

Careful selection of these effect modifiers is crucial. On the one hand, deciding on the appropriate set of \textit{observed} quantities is difficult, and a large set of candidates could arise. One key objective of this paper is to provide techniques to select promising subsets. On the other hand, appropriately selecting the \textit{latent} effect modifiers allows us to capture situations where parts of the coefficients evolve smoothly whereas others move more abruptly. The latter behavior of the TVPs is often found for US macroeconomic data \citep[see][]{sims2006were}, whereas the former is consistent with financial time series such as bond yields or stock returns \citep[see][]{dh2012, huber2019should}.  

Since large VARs often include both macroeconomic as well as financial quantities, a successful model should be able to accommodate both types of structural change or even rely on linear combinations of them. This is precisely what we aim to achieve in this paper. Our  approach is capable of answering not only the question why coefficients change, but also to infer an appropriate law of motion using a broad  set of latent quantities, each equipped with its own state evolution equation. These unobserved quantities range from latent factors that follow a random walk to Markov switching indicators that allow a subset of parameters to switch between a low number of regimes. Our model approach nests several alternatives proposed in the literature such as the TVP-VAR of \citet{primiceri2005} and \citet{cogley2005drifts} or the reduced-rank model of \cite{chan2020reducing}.

This large degree of flexibility, however, comes with two concerns. The first is that overfitting problems can easily arise. We overcome these by using Bayesian shrinkage priors. Our prior is a variant of the well-known Horseshoe prior \citep[see][]{carvalho2010horseshoe} that allows us to shrink coefficients associated with irrelevant effect modifiers towards zero. The second concern relates to computation. Since inclusion of a large number of endogenous variables in a VAR quickly leads to a huge dimensional parameter space, we propose a computationally efficient Markov chain Monte Carlo (MCMC) algorithm. To circumvent mixing issues, we rely on different parameterizations of the model during MCMC sampling. The corresponding novel algorithm thus provides a second important contribution of the paper.

In our empirical work, we use the model approach to forecast the US term structure of interest rates. We investigate the empirical properties of our approach using two information sets in the underlying VAR. First, we include several interest rates at different maturities directly as endogenous variables. Second, we consider a three-factor Nelson-Siegel model for the term structure of interest rates as in \citet{diebold2006forecasting} and \citet{diebold2008global}. Adopting a long hold-out period that includes several recessionary episodes, our approach improves upon a wide range of competing models. While improvements for point forecasts are often muted, we find that our proposed model yields favorable density predictions. The predictive exercise is complemented by a comprehensive discussion of patterns in time-variation and their sources. Moreover, how our approach can be used to analyze low frequency relations between the observed quantities in the model over time.

The rest of the paper is structured as follows. Section \ref{sec: rotation} introduces the econometric framework which includes the general form of the TVP-VAR, a flexible law of motion for the latent states as well as the effect modifiers which crucially impact the state dynamics. This section, moreover, introduces the Bayesian prior setup and techniques for  posterior  and predictive inference. Section \ref{sec: empApp} applies the model approach to the term structure of US interest rates. It also serves to illustrate key model features and to highlight the predictive capabilities of the approach in an out-of-sample forecasting exercise. The last section summarizes and concludes the paper. Additional technical details and further empirical results are provided in the Appendix.

\section{Flexible Bayesian Inference in TVP-VARs}\label{sec: rotation}
\subsection{The TVP-VAR}\label{sec: regmod}
Let $\bm y_t$ denote an $M \times 1$ vector of macroeconomic and financial quantities at time $t=1, \dots, T$.  We assume that $\bm y_t$ depends on its $P$ lags which we store in a $K=MP$-dimensional vector $\bm x_t = (\bm y'_{t-1}, \dots, \bm y'_{t-P})'$. Then the basic TVP-VAR can be written as a linear multivariate regression model:
\begin{equation*}
\bm y_t = (\bm I_M \otimes \bm x'_t) \bm \beta_t + \bm \epsilon_t,
\end{equation*}
where $\bm \beta_t$ represents a set of $k = MK$ dynamic regression coefficients and $\bm \epsilon_t \sim \mathcal{N}(\bm 0_M, \bm \Sigma_t)$ is a vector Gaussian shock process with time-varying $M \times M$-dimensional variance-covariance matrix $\bm \Sigma_t$. We assume that $\bm \Sigma_t$ can be decomposed as follows:
\begin{equation*}
\bm \Sigma_t = \bm Q_t \bm H_t \bm Q'_t.
\end{equation*}
Here, $\bm Q_t$ denotes an $M \times M$ lower triangular matrix with unit diagonal with $v (= M (M-1))$ free elements denoted by $\bm q_t$. $\bm H_t = \text{diag}(e^{h_{1t}}, \dots, e^{h_{Mt}})$ is a diagonal matrix with $h_{jt}~(j=1, \dots, M)$ representing time-varying log-volatilities. These are assumed to evolve according to an AR(1) process:
\begin{equation*}
h_{jt} = \mu_j + \psi_j (h_{jt-1} - \mu_j) + \nu_{jt}, \quad \nu_{jt} \sim \mathcal{N}(0,  \varsigma^2_j).
\end{equation*}
$\mu_j$ is the unconditional mean, $\psi_j$ the persistence parameter and $\varsigma_j^2$ the variance of the log-volatility process for equation $j$.

In what follows, we rewrite the TVP-VAR using its non-centered parameterization \citep{sfs_wagner2010}:
\begin{equation*}
\bm y_t = (\bm I_M \otimes \bm x'_t) (\bm \beta + \tilde{\bm \beta}_t) + ({\bm{Q}} + \tilde{\bm{Q}}_t) \bm{\varepsilon}_t, \quad \bm{\varepsilon}_t\sim\mathcal{N}(\bm{0}_M,\bm{H}_t).
\end{equation*}
$\bm \beta$ denotes a $k$-dimensional vector of constant coefficients, and $\tilde{\bm \beta}_t = \bm \beta_t - \bm \beta$. This parameterization allows us to disentangle time-invariant (encoded by $\bm \beta$) from time-varying effects (encoded by $\tilde{\bm \beta}_t$) for the regressors. For the decomposed variance-covariance matrix, we have $\bm{Q}$, a lower triangular matrix with ones on the diagonal capturing the constant part of the covariances. $\tilde{\bm{Q}}_t = \bm{Q}_t - \bm{Q}$ is the corresponding lower triangular matrix with zero-diagonal elements containing the time-varying part. Their free elements are collected in the $v\times1$-vectors $\bm{q}$ and $\tilde{\bm{q}}_t$, respectively.

In this paper, the focus is on modeling the $N (=k+v)$-dimensional vector $\tilde{\bm{\gamma}}_t = (\tilde{\bm \beta}_t',\tilde{\bm{q}}_t')'$ and the constant part  ${\bm{\gamma}} = ({\bm \beta}',{\bm{q}}')'$. The literature typically assumes that the transition distribution $p(\tilde{\bm \gamma}_t| \tilde{\bm \gamma}_{t-1})$ is given by:
\begin{equation*}
\tilde{\bm \gamma}_t|\tilde{\bm{\gamma}}_{t-1} \sim  \mathcal{N}(\tilde{\bm \gamma}_{t-1}, \bm V) \text{ with }  \tilde{\bm \gamma}_0 = \bm 0_N.
\end{equation*}
This law of motion suggests that the expected value of $\tilde{\bm \gamma}_{t-1}$ equals $\tilde{\bm \gamma}_{t}$ and the amount of time-variation is determined by the $N \times N$-dimensional process innovation variance-covariance matrix $\bm V$. This matrix is often assumed to be diagonal. Notice that if selected elements in $\bm V$ are equal to zero, the corresponding regression coefficients are constant.

Estimation and inference is typically carried out via Bayesian methods. The recent literature proposes using shrinkage priors to allow for data-based selection of those coefficients which should be time-varying or constant. This already leads to substantial improvements in predictive accuracy but does not tackle the fundamental question whether the coefficients are better characterized by a random walk, a change-point process, or by mixtures of these.

\subsection{A General Specification for the TVPs}
As discussed in the previous sub-section, the typical assumption is that $\tilde{\bm \gamma}_t$ follows a random walk process. In addition, the shocks to the random walk state equation are often assumed to feature a positive error variance. We relax both assumptions to allow for more flexibility. 

The random walk assumption is relaxed by assuming that the time-varying part stored in $\tilde{\bm{\gamma}}_t$ depends on a set of $R$ additional  factors $\bm z_t$. These $\bm z_t$ are the effect modifiers mentioned in the introduction that can be observed or latent. The relationship between the TVPs and $\bm z_t$ is given by:
\begin{equation}
\tilde{\bm{\gamma}}_t = \bm \Lambda \bm z_t + \bm \eta_t.\label{eq: obs_states}
\end{equation}
$\bm \Lambda$ denotes an $N \times R$ matrix of regression coefficients, and $\bm \eta_t \sim \mathcal{N}(\bm 0_N, \bm \Omega)$ is a Gaussian error term with diagonal error variance-covariance matrix $\bm \Omega = \text{diag}(\omega^2_1, \dots, \omega^2_N)$. If $R \ll N$, the coefficients feature a factor structure and co-move according to the effect modifiers in $\bm z_t$. The relationship between $\tilde{\bm{\gamma}}_t$ and $\bm z_t$ is determined by the factor loadings in $\bm \Lambda$. For instance, if the $j^{\text{th}}$ column of $\bm \Lambda, \bm \lambda_j$, is equal to zero, the corresponding $j^{\text{th}}$ factor in $\bm z_t$ does not enter the model and thus has no influence on $\tilde{\bm{\gamma}}_t$. 

 %Several choices are possible and we discuss each of them in Sub-section \ref{sec: effects}.
The specific selection of $\bm z_t$ is crucial for determining the dynamics of $\tilde{\bm{\gamma}}_t$. Appropriate choice of $\bm z_t$ yields a variety of important special cases that depend on the specific values of $\bm \Lambda$ and $\bm \Omega$ as well as on the composition of $\bm z_t$. In this sub-section, we briefly focus on special cases that arise independently of the choice of $\bm z_t$. The next sub-section deals with cases that arise if $\bm z_t$ is suitably chosen.

These two cases are the following. If $\bm \Lambda = \bm 0_{N \times R}$, with $\bm 0_{N \times R}$ being a $N \times R$ matrix of zeros, we obtain a random coefficients model that assumes that the regression coefficients follow a white noise process \citep[for some recent papers that follow this approach, see][]{korobilis2019high, hhko2019}. The second special case arises if both $\bm \Lambda = \bm 0_{N \times R}$ and $\bm \Omega = \bm 0_{N \times N}$. In this case, we obtain a standard constant parameter regression model.  

Before we discuss the choice of $\bm z_t$, it is worth noting that if $\bm z_t$ is (partially) latent, the model in Eq. (\ref{eq: obs_states}) is not identified. Since our object of interest is $\bm \gamma_t$, this poses no greater issues. If we wish to structurally interpret elements in $\bm z_t$, standard identification strategies from the literature on dynamic factor models can be used \citep[see, e.g.,][]{geweke1996measuring,aguilar2000bayesian, stock2011dynamic}.

\subsection{Possible Choices for the Effect Modifiers}\label{sec: effects}
The specific choice of $\bm z_t$ is crucial in determining how $\tilde{\bm{\gamma}}_t$ behaves over time. Hence, by suitably choosing the elements in $\bm z_t$, our model approach is related to the following specifications:
\begin{itemize}[leftmargin=*]
\item \cite{chan2020reducing}: We assume that $\bm z_t$  consists exclusively of a sequence of $R = R_\tau$ latent factors $\bm \tau_t$, which follow a multivariate random walk:
\begin{equation*}
\bm \tau_{t} = \bm \tau_{t-1} + \bm \nu_t, \quad \bm \nu_t \sim \mathcal{N}(0, \bm V_\tau).
\end{equation*}
$\bm V_\tau = \text{diag}(v_1^2, \dots, v^2_{R_\tau})$ denotes a diagonal  variance-covariance matrix with $v_j^2$ being process innovation variances that determine the smoothness of the elements in $\bm \tau_t$. Note that setting $v_j^2$ close to zero effectively implies that $\tau_{jt}$, the $j^{\text{th}}$ element in $\bm \tau_t$, is constant.  This model implies a factor structure in $\tilde{\bm \gamma}_t$ if $R_\tau \ll N$.

\item \cite{primiceri2005}: If $R = R_\tau = N$, the elements in $\bm z_t$ are  random walks, and $\bm \Lambda = \bm I_N$, we obtain a standard time-varying parameter model. Assuming that the covariances are constant we obtain the model put forth in \cite{cogley2005drifts}.

\item \cite{sims2006were}: A Markov switching model can be obtained by setting $\bm z_t = S_t$, with $S_t \in \{0, 1\}$ denoting a binary indicator with transition probabilities given by:
\begin{equation*}
p(S_t = i | S_{t-1} = j) = p_{ij} \quad \text{for } i,j = 0,1,
\end{equation*}
with $p_{ij}$ denoting the $(i,j)^{\text{th}}$ element of a $2 \times 2$-dimensional transition probability matrix $\bm P$. Inclusion of this random quantity allows to capture structural breaks in $\tilde{\bm{\gamma}}_t$ that are common to all coefficients. 

\item \cite{caggiano2017estimating}: Assuming that $\bm z_t$ is exclusively composed of observed quantities we obtain a regression model with interaction terms. 
\end{itemize}
These examples show that our model, conditional on choosing a suitable set of effect modifiers, is capable of mimicking several prominent specifications in the literature. Since the question on the appropriate state evolution equation is essentially a model selection issue, we simply specify $\bm z_t$ to include most (with the exception of the $R=N$ setup) of the modifiers discussed above. 

More precisely, we set $\bm z_t$ as follows:
\begin{equation*}
\bm z_t = (\bm r'_t, \bm{S}'_t, \bm \tau'_t)'.
\end{equation*}
Here we let $\bm r_t$ denote a set of $R_r$ observed factors and the dimension of $\bm z_t$ is thus $R = R_r  + R_S + R_\tau$ with $R_S = M$. To allow for additional flexibility we assume that $\bm \tau_t = (\bm \tau_{1t}', \dots, \bm \tau_{Mt}')'$ with $\bm \tau_{jt}$ being equation-specific factors  of dimension $R_{\tau j}$ (and thus $R_\tau = \sum_j R_{\tau j}$). Assuming that $R_{\tau i} = R_{\tau j} = \delta$ for all $i,j$, we have  $R_\tau = \delta M$ latent random walk factors. Likewise, we estimate a separate Markov switching indicator $S_{jt}$ per equation (and thus $\bm S_t = (S_{1t}, \dots, S_{Mt})'$, with the corresponding transition probabilities matrix denoted by $\bm{P}_j$). 

The corresponding loadings matrix $\bm \Lambda$ is structured such that the loadings in equation $j$ associated with the factors $\bm \tau_{it}$ and $S_{it}$ for $i \neq j$ equal zero. This assumption strikes a balance between assuming a large number of latent factors to achieve maximum flexibility (and thus risk overfitting) and using a rather parsimonious model (with the risk of being too simplistic). Recent contributions use similar assumptions on the state evolutions, \citep[see][]{koop2009evolution, maheu2018efficient}. As opposed to these papers, our approach offers more flexibility since, if necessary, the presence of the idiosyncratic shocks to the TVPs allows for deviations if the factor structure does not represent the data well.

Our specification implies that, depending on the factor loadings $\bm \Lambda$, the evolution of $\tilde{\bm{\gamma}}_t$ might be a combination of a set of random walk factors, a Markov switching process and some observed quantities. To single out irrelevant elements in $\bm z_t$, one could simply set the corresponding columns in $\bm \Lambda$ equal to zero. In this paper, we achieve this through a Bayesian shrinkage prior. The next sub-section discusses our priors in more detail.

\subsection{The Prior Setup}\label{prior}
The discussion in Sub-section \ref{sec: regmod} shows that our model approach nests a variety of competing models. To select the appropriate model variant and alleviate over-parameterization concerns, we opt for a Bayesian approach to introduce shrinkage. Here, we summarize the priors we impose on key parameters.

% Priors on the loadings
In light of the specific choice of $\bm z_t$, we introduce some additional notation to clarify details on our prior implementation. Let us assume that $\bm \Lambda$ is composed of the following matrices:
\begin{equation*}
\bm \Lambda = [\bm \Lambda_r \quad \bm \Lambda_{S} \quad \bm \Lambda_\tau],
\end{equation*}
where $\bm \Lambda_r$ is an $N \times R_r$ matrix of loadings related to the observed quantities, $\bm \Lambda_{S}$ denotes an $N \times R_{S}$ matrix of loadings related to $\bm{S}_t$, and $\bm \Lambda_\tau$ represents an $N \times R_\tau$ factor loadings matrix associated with $\bm \tau_t$. 

For imposing shrinkage we rely on variants of the horseshoe prior \citep{carvalho2010horseshoe}. While in principle any global-local shrinkage prior may be used, we choose the horseshoe prior due to its excellent shrinkage properties and the lack of tuning parameters. In particular, we specify a column-wise horseshoe prior on the loadings matrix. Let $\bm{\Lambda}_j$ denote a sub-matrix of the free elements in $\bm{\Lambda}$ corresponding to the $j^{\text{th}}$ equation, and $\bm{\lambda}_{ji}$ mark the $i^{\text{th}}$ column of this matrix. $\lambda_{ji,\ell}$ refers to the $\ell^{\text{th}}$ element of this vector. The prior is given by:
\begin{equation*}
\lambda_{ji,\ell}|\kappa_{ji,\ell},\delta_{ji}\sim\mathcal{N}(0,\kappa_{ji,\ell}^2\delta_{ji}^2), \quad \kappa_{ji,\ell}\sim\mathcal{C}^{+}(0,1), \quad \delta_{ji}\sim\mathcal{C}^{+}(0,1).
\end{equation*} 
Here, $\mathcal{C}^{+}(0, 1)$ denotes the half Cauchy distribution, and $\delta_{ji}$ is an equation- and column-specific global shrinkage factor, while $\kappa_{ji,\ell}$ is a local scaling parameter.

% Priors on the constant part of the parameters
To further regularize our potentially huge-dimensional parameter space, we impose an equation-wise horseshoe prior on the constant part of the regression coefficients and covariances in $\bm{\gamma}_j$ corresponding to the $j^{\text{th}}$ equation. Let $\gamma_{ji}$ denote the $i^{\text{th}}$ element of the vector.  The setup is similar to the one of the loadings matrix, and given by the hierarchical structure:
\begin{equation*}
\gamma_{ji}|\xi_{ji},\zeta_j\sim\mathcal{N}(0,\xi_{ji}^2\zeta_j^2), \quad \xi_{ji}\sim\mathcal{C}^{+}(0,1), \quad \zeta_j\sim\mathcal{C}^{+}(0,1).
\end{equation*}
The hyperparameter $\zeta_j$ is an equation-specific global shrinkage factor, while the $\xi_{ji}$'s are local scalings.

% Priors for the variances
We mentioned earlier that it is often assumed that shocks to the states feature positive error variances. To introduce shrinkage of $\omega_{ii}$ towards zero, we impose a horseshoe prior also on the square root of the innovation variances of the measurement errors in Eq. (\ref{eq: obs_states}).  This prior is specified in an equation-specific manner. For equation $j$, let $\bm{\omega}_j$ denote a $v_j (=j-1 + k)$-dimensional vector which stores the diagonal elements in  $\bm{\Omega}$ associated with the $j^{\text{th}}$ equation.  This includes the process innovation variances on the  $k$ regression coefficients and the $j-1$ covariance parameters in $\tilde{\bm  Q}_t$. The square root of the $i^{\text{th}}$ element of $\bm \omega_j$, $\sqrt{\omega_{ji}}$, features the following prior hierarchy:
\begin{equation*}
\sqrt{\omega_{j i}}|\varpi_{ji},\vartheta_j \sim \mathcal{N}(0,\varpi_{ji}^2\vartheta_j^2), \quad \varpi_{ji}\sim\mathcal{C}^{+}(0,1), \quad \vartheta_j\sim\mathcal{C}^{+}(0,1).
\end{equation*}
Choosing a Gaussian prior on the square root of the variance in the first level of the hierarchy implies a Gamma prior on $\omega_{ji}$, with $\omega_{ji}|\varpi_{ji},\vartheta_j\sim\mathcal{G}\left(1/2,\varpi_{ji}^{-2}\vartheta_j^{-2}/2\right)$, see also \citet{sfs_wagner2010}. The hyperparameters $\vartheta_j$ and $\varpi_{ji}$ are again equation-specific global and local shrinkage parameters.  Furthermore, we set $\bm{V}_\tau = \bm{I}_{R_\tau}$ and thus impose shrinkage through the factor loadings in $\bm{\Lambda}_\tau$ \citep[see][]{chan2020reducing}. 

% SV priors
On the parameters of the state equation of the log-volatility processes $\mu_j, \psi_j$ and $\varsigma_j^2$, we use the setup proposed in \cite{kastner2014ancillarity}. That is, we assume a Gaussian prior on the unconditional mean, $\mu_j \sim \mathcal{N}(0, 10^2)$, a Beta prior on the transformed autoregressive parameter, $(\psi_j + 1)/2 \sim \mathcal{B}(5, 1.5)$, a Gamma prior on the state variance, $\varsigma_j^2 \sim \mathcal{G}(1/2, 1/2)$, while the prior on the initial state is the unconditional distribution for all $j=1,\dots, M$.

For the equation-specific transition probabilities $\bm{P}_j$ of Markov switching indicators, we assume that the $(i, i)^{\text{th}}$ element $p_{j,ii}$ arises from a Beta distribution given by: 
\begin{equation*}
p_{j, ii} \sim \mathcal{B}(e_{i0}, e_{i1}), \quad \text{for } i = 0,1 \text{, } j = 1, \dots, M,  
\end{equation*}
and hence $p_{j, i\ell} = 1 - p_{j,i\ell}$ for $i \neq \ell$. In the empirical application, we define $e_{00} = e_{11} = 10$ and $e_{01} = e_{10} = 1$, in order to weakly push each $S_{jt}$ towards a single-state a priori. 

\subsection{Full Conditional Posterior Simulation}
To simulate from the full posterior distribution we develop an efficient MCMC algorithm. Since full-system estimation of the VAR quickly becomes computationally cumbersome, we rely on the equation-by-equation algorithm suggested in \cite{ccm2019}. 

Conditional on $\bm Q_t$, one can state the VAR as a system of (conditionally) independent equations. The first equation of this system is given by:
\begin{equation*}
y_{1t} = \bm x'_t (\bm \beta_1 + \tilde{\bm \beta}_{1t}) + \varepsilon_{1t},
\end{equation*}
and  the $j^{\text{th}}$ equation $ (j > 1)$:
\begin{align}
y_{jt} = \bm x'_t (\bm \beta_j + \tilde{\bm \beta}_{jt}) + \bm u'_{jt} (\bm q_{j} + \tilde{\bm q}_{jt})   + \varepsilon_{jt}. \label{eq: jtheq-by-eq}
\end{align}
${\bm{\beta}}_{j}$ and $\tilde{\bm{\beta}}_{jt}$ denote the $j^{\text{th}}$ subvectors of the constant and time-varying parts in $\bm \beta_t$ with $\bm \beta_t = (\bm \beta'_{1t}, \dots, \bm \beta'_{Mt})'$ and  $\bm u_{jt} = (\varepsilon_{1t}, \dots, \varepsilon_{j-1, t})'$. The $(j-1)$-dimensional vectors $\bm q_j$ and $\tilde{\bm{q}}_{jt}$ store the constant and time-invariant part of the free elements in the $j^{\text{th}}$ row of $\bm Q_t$. This approach allows to estimate the different elements of $\bm \gamma_t$ that relate to the $M$ equations independently from each other conditional on the shocks to the preceding $j-1$ equations. This speeds up computation enormously.

Equation (\ref{eq: jtheq-by-eq}) can be simplified to yield:
\begin{equation}
y_{jt} = \bm m'_{jt} (\bm \gamma_{j} + \tilde{\bm \gamma}_{jt}) + \varepsilon_{jt}, \label{eq: reg_by_reg}
\end{equation}
where $\bm m_{jt} = (\bm x'_t, \bm u_{jt}')'$, $\tilde{\bm \gamma}_{jt} = \bm \gamma_{jt} - \bm \gamma_{j}$, and $\bm \gamma_{jt}$ refers to the TVPs associated with the $j^{\text{th}}$ equation in $\bm \gamma_t$, and $\bm{\gamma}_j$ denotes the corresponding constant part. All the following steps will be carried out on an equation-by-equation basis and making use of the regression form in Eq. (\ref{eq: reg_by_reg}). For notational simplicity, we assume that all elements in $\bm z_t$ are latent. In light of the discussion in Sub-section \ref{sec: effects}, this implies that $\bm z_{jt} = (\bm \tau'_{jt},~ S_{jt})'$ and the extension to include observed factors is trivial.

\textbf{Sampling $\bm z_{jt}$}. Conditional on the remaining quantities of the model, we simulate the latent (random walk and Markov switching) components in $\bm z_{jt}$ by integrating out $\tilde{\bm{\gamma}}_{jt}$. This is achieved by rewriting Eq. (\ref{eq: reg_by_reg}) as:
\begin{equation}
\tilde{y}_{jt} =  \bm m'_{jt} \bm \Lambda_j \bm z_{jt} + \bm m'_{jt} \bm \eta_{jt} + \varepsilon_{jt}, \label{eq: reg_by_reg_NCP}
\end{equation}
with $\tilde{y}_{jt} = y_{jt} - \bm m'_{jt} \bm \gamma_{j}$. Defining $\tilde{\bm m}'_{jt} = \bm m'_{jt} \bm \Lambda_j$  and $\hat{\varepsilon}_{jt} = \bm m'_{jt} \bm \eta_{jt} + \varepsilon_{jt}$ allows us to cast Eq. (\ref{eq: reg_by_reg_NCP}) as a simple linear regression model:
\begin{equation}
\tilde{y}_{jt} =  \tilde{\bm m}'_{jt} \bm z_{jt} + \hat{\varepsilon}_{jt}, \quad \hat{\varepsilon}_{jt} \sim \mathcal{N}(0, \bm m'_{jt} \text{diag}(\bm \omega_j) \bm m_{jt} + e^{h_{jt}}).
\end{equation}
This parameterization has the advantage that it does not depend on $\tilde{\bm \gamma}_{jt}$, and $\bm z_{jt}$ can thus be sampled marginally of $\tilde{\bm \gamma}_{jt}$. This improves mixing substantially since $\tilde{\bm \gamma}_{jt}$ and $\bm z_{jt}$ will often be highly correlated \citep[for a detailed discussion of this issue, see][]{gerlach2000efficient, giordani2008efficient}.

Depending on the precise law of motion for the elements in $\bm z_{jt}$, standard algorithms can now be used. In this paper, we use two different law of motions. For the latent random walk factors in $\bm \tau_{jt}$, we use the forward filtering backward sampling algorithm outlined in \cite{carter1994gibbs} and \cite{fs1994}. In case of the latent Markov switching factors in $S_{jt}$, we use the algorithm outlined in \cite{kim1999state}. Both algorithms are well known and relevant details may be found in the original papers. Here, it suffices to note that in both cases, sampling the latent states is computationally easy since the state space is low dimensional with $R \ll N$. In this setting, sampling the factors equation-wise can be carried out in ${O}(R^3)$ steps, a substantial computational improvement relative to the ${O}(N^3)$ steps necessary to estimating an unrestricted TVP regression \citep[see also the discussion in][]{chan2020reducing}.\footnote{It is worth mentioning that the ${O}(N^3)$ statement is true for the precision sampler and differs for forward-filtering backward-sampling algorithms.}

\textbf{Sampling the state innovation variances}. To obtain draws for the state innovation variances, reconsider Eq. (\ref{eq: obs_states}) and draw them conditional on the observed/latent states and the factor loadings using a generalized inverse Gaussian distribution. For further details and the moments of this distribution, see Appendix \ref{app:technical}.

\textbf{Sampling $\bm \Lambda_j$ and $\bm \gamma_{j}$ jointly}. Similarly to $\bm z_{jt}$, we sample the non-zero loadings in $\bm \Lambda$ and the time-invariant coefficients marginally of $\bm \gamma_{jt}$ by using equation-by-equation estimation. The observation equation for equation $j$ (conditional on $\bm z_{jt}$) can be written as a standard regression model: 
\begin{equation}
y_{jt} = \hat{\bm m}_{jt}' \hat{\bm \gamma}_{j} + \hat{\varepsilon}_{jt}, \label{eq: reg_by_reg_NCP_LAMBDA}
\end{equation}
where $\hat{\bm m}_{jt} = (\bm m_{jt}', (\bm z_{jt} \otimes \bm m_{jt})')'$ is an $R v_j^2$-dimensional vector of covariates, and $ \hat{\bm \gamma}_{j} = (\bm \gamma_{j}', \text{vec}(\bm \Lambda_j)')'$ denoting an $R v_j^2$-dimensional coefficient vector. The posterior of $\hat{\bm \gamma}_{j}$ is Gaussian with well known moments.

\textbf{Sampling the stochastic volatilities}. The latent log-volatility processes can again be sampled on an equation-by-equation basis. This step is implemented using the \texttt{R}-package \texttt{stochvol}.

\textbf{Sampling the horseshoe prior hyperparameters}. Our assumptions imply analogous horseshoe priors for the factor loadings matrix, the constant part of the coefficients and the square root of the state innovation variances. Posteriors are provided in Appendix \ref{app:technical}.

\section{Forecasting US Government Bond Yields}\label{sec: empApp}
This section applies the model to predict the term structure of US interest rates. These time series are characterized by substantial non-linearities (e.g. during the period of the zero lower bound), feature substantial co-movement both in the level of the time series but also in the parameters describing their evolution. Our proposed model framework might thus be well suited to capture such features. We investigate this claim in a thorough forecasting exercise using several established benchmarks. After showing that our approach yields favorable forecasts, we discuss the driving forces behind parameter changes as well as discuss how key quantities that shape yield curve dynamics co-move over time at low frequencies.

\subsection{Data and Design of the Forecasting Exercise}
Our aim is to predict  monthly zero-coupon yields of US treasuries at different yearly maturities. The data is described in detail in \citet{gurkaynak2007us}.\footnote{Available online at \href{https://www.federalreserve.gov/data/nominal-yield-curve.htm}{federalreserve.gov/data/nominal-yield-curve.htm}.} The target variables are $1,3,5,7,10 \text{ and } 15$ years maturities. All variables enter our model in first differences.

Estimation and forecasting is carried out recursively. Using data from 1973:01 to 1999:12, we produce one-month-ahead and three-months-ahead forecasts for 2000:01. After obtaining the predictive distributions we expand the sample and repeat this procedure until we reach 2019:12.\footnote{Note that for our set of financial indicators, data revisions and ragged edges arising from delays in the publication of the series do not matter. This is due to financial market data being available almost instantaneously, and the published quotes are not subject to revisions at later dates.}  Point forecast performance is measured using Root Mean Squared Errors (RMSEs), while density forecasts are assessed in terms of Log Predictive Bayes Factors (LPBFs), averaged over the out-of-sample observations \citep{geweke2010comparing}. 

For analyzing the role of observed factors governing joint dynamics of the TVPs, we rely on three exogenous variables: 
\begin{enumerate}[(a)]
\item a binary recession indicator (labeled REC) for the US, dated on a monthly basis by the Business Cycle Dating Committee of the National Bureau of Economic Research,
\item the National Financial Conditions Index (NFCI), maintained by the Federal Reserve Bank of Chicago, downloaded from the FRED database of the Federal Reserve Bank of St. Louis (available online at \href{https://fred.stlouisfed.org}{fred.stlouisfed.org}), and
\item  the Risk-Free (RF) interest rate from the Fama-French Portfolios and Factors database, provided on the web page of Kenneth R. French as another important early warning indicator.\footnote{Available online at \href{https://mba.tuck.dartmouth.edu/pages/faculty/ken.french}{mba.tuck.dartmouth.edu/pages/faculty/ken.french}.}  Thus, $\bm{r}_t=({r}_{\text{NGCI}, t-1}, {r}_{\text{REC},t-1}, {r}_{\text{RF},t-1})'$ and $R_r = 3$. Exogenous variables enter the model as first order lags. Higher-order forecasts involving exogenous variables are based on random walk predictions of these quantities.
\end{enumerate}
We use these three effect modifiers for simple reasons. First, there is strong evidence that yield curve dynamics differ across business cycle phases \citep[see, e.g.,][]{hevia2015estimating}. Second, the RF interest rate serves as an early warning indicator which possesses predictive power for changes in the shape of the yield curve. Finally, the inclusion of the NFCI is motivated by the recent literature on forecasting tail risks in macroeconomic and financial time series \citep[see, e.g.,][]{adrian2019vulnerable, carriero2020capturing, adams2021forecasting}.

\subsection{Competing Model Specifications}
The forecast exercise distinguishes between two model classes, with 15 distinct model specifications in each class. The first model class involves TVP-VARs that incorporate the six target variables as endogenous variables, that is $M=6$. This model class is labeled as VAR.

The second model class includes specifications based on the three-factor Nelson-Siegel (NS) model as in \citet{diebold2006forecasting} that imposes a factor structure on the yields:
\begin{equation*}
i _{t}(\theta) = L_{t} + \left(\frac{1-\exp(- \theta \alpha)}{\theta \alpha}\right) \text{\textit{\v{S}}}_t  + 
\left( \frac{1-\exp(- \theta \alpha)}{\theta \alpha} - \exp(- \theta \alpha) \right) \text{\textit{\c{C}}}_t.  \label{eq:NS}
\end{equation*}
where $i_t(\theta)$ denotes the yield at maturity $\theta$ at time $t$, $L_t$ is a factor that controls the level, $\text{\textit{\v{S}}}_t$ determines the slope, and $\text{\textit{\c{C}}}_t$ represents the curvature factor of the yields. The parameter $\alpha$ governs the exponential decay rate. To maximize the loading on $\text{\textit{\c{C}}}_t$ we set $\alpha=0.7308\ (12\times0.0609)$.\footnote{See \citet{diebold2006forecasting} for a discussion of this specific choice.} In what follows, we use the latent factors $L_t,\text{\textit{\v{S}}}_t \text{ and } \text{\textit{\c{C}}}_t$ as endogenous variables in the VAR specifications by defining $\bm{y}_t=(L_t, \text{\textit{\v{S}}}_t, \text{\textit{\c{C}}}_t)^\prime$, resulting in $M=3$. These latent factors are obtained by running OLS on a $t$-by-$t$ basis. This model class is subsequently labeled NS-VAR.

Model specifications are differentiated over a grid of effect modifier combinations. In particular, specifications within a model class differ in terms of three aspects (see Table \ref{tab:LPS}): First, in terms of whether the three exogenous variables (collected in $\bm{r}_t$) are included in $\bm{z}_t$ or not ("x" marks inclusion, "--" indicates no observed factors); second, in terms of the number of latent random walk factors $R_{\tau j}$ included in $\bm{z}_t$ (which we assume to be equal across equations); and third, in terms of the presence of Markov switching indicators in $\bm{S}_t$, again with "x" marking their inclusion and "--" their absence. This setup implies that we have 15 time-varying parameter NS-VAR and 15 time-varying parameter VAR model specifications. For comparative purposes we also consider the two class-specific constant parameter model variants, labeled ``Constant,'' and a conventional independent random walk specification of the TVPs (i.e. we set the number of random walk factors equal to $K$ and exclude $\bm r_t$ and $\bm S_t$). Notice that we also have a specification which includes only $\bm S_t$ and another one which uses only observed factors. The latter one is closely related to a Markov switching model whereas the second one closely resembles a VAR with interaction terms.

\subsection{Results}\label{sec: forecasting}
Table \ref{tab:LPS} shows the one-month and one-quarter-ahead out-of-sample forecasting results for US treasury yields at different maturities, using the 30 TVP model specifications and the two constant parameter model variants as described in the previous sub-section. Recall that $M=3$ (and $K=9$) in case of the NS-VAR and $M=6$ (and $K=18$) in case of the VAR model variants. $\bm{z}_t=(\bm{r}_t^\prime, \bm{S}_t^\prime, \bm{\tau}_t^\prime)$ where $\bm{r}_t$ denotes $R_r(=3)$-dimensional vector, $\bm{S}_t$ a $R_{S}$-dimensional vector, and  $\bm{\tau}_t$ a $R_\tau$-dimensional vector. 
All specifications feature $P=3$ lags of the endogenous variables.

\afterpage{%
\clearpage%
\onehalfspacing
\begin{landscape}
\begin{tiny}
\begin{longtable}{llllclclllllllcllllllll}
\caption{Out-of-sample forecasting results for US treasury yields at different maturities using TVP-NS-VAR and TVP-VAR model specifications.}\label{tab:LPS}\tabularnewline
\toprule
\multicolumn{1}{l}{\bfseries Model}&\multicolumn{3}{c}{\bfseries Specification}&\multicolumn{1}{c}{\bfseries }&\multicolumn{1}{c}{\bfseries }&\multicolumn{1}{c}{\bfseries }&\multicolumn{7}{c}{\bfseries One-month-ahead}&\multicolumn{1}{c}{\bfseries }&\multicolumn{1}{c}{\bfseries }&\multicolumn{7}{c}{\bfseries One-quarter-ahead}\tabularnewline
\cmidrule{2-4} \cmidrule{8-14} \cmidrule{17-23}
\multicolumn{1}{l}{}&\multicolumn{1}{c}{$\bm{r}_t$}&\multicolumn{1}{c}{$\delta$}&\multicolumn{1}{c}{$\bm{S}_t$}&\multicolumn{1}{c}{}&\multicolumn{1}{c}{}&\multicolumn{1}{c}{}&\multicolumn{1}{l}{Joint}&\multicolumn{1}{l}{1 year}&\multicolumn{1}{l}{3 year}&\multicolumn{1}{l}{5 year}&\multicolumn{1}{l}{7 year}&\multicolumn{1}{l}{10 year}&\multicolumn{1}{l}{15 year}&\multicolumn{1}{c}{}&\multicolumn{1}{c}{}&\multicolumn{1}{l}{Joint}&\multicolumn{1}{l}{1 year}&\multicolumn{1}{l}{3 year}&\multicolumn{1}{l}{5 year}&\multicolumn{1}{l}{7 year}&\multicolumn{1}{l}{10 year}&\multicolumn{1}{l}{15 year}\tabularnewline
\midrule
\endfirsthead

\multicolumn{23}{r}{\footnotesize{\textit{Table \ref{tab:LPS} continued}}}\\
\toprule
\multicolumn{1}{l}{\bfseries Model}&\multicolumn{3}{c}{\bfseries Specification}&\multicolumn{1}{c}{\bfseries }&\multicolumn{1}{c}{\bfseries }&\multicolumn{1}{c}{\bfseries }&\multicolumn{7}{c}{\bfseries One-month-ahead}&\multicolumn{1}{c}{\bfseries }&\multicolumn{1}{c}{\bfseries }&\multicolumn{7}{c}{\bfseries One-quarter-ahead}\tabularnewline
\cmidrule{2-4} \cmidrule{8-14} \cmidrule{17-23}
\multicolumn{1}{l}{}&\multicolumn{1}{c}{$\bm{r}_t$}&\multicolumn{1}{c}{$\delta$}&\multicolumn{1}{c}{$\bm{S}_t$}&\multicolumn{1}{c}{}&\multicolumn{1}{c}{}&\multicolumn{1}{c}{}&\multicolumn{1}{l}{Joint}&\multicolumn{1}{l}{1 year}&\multicolumn{1}{l}{3 year}&\multicolumn{1}{l}{5 year}&\multicolumn{1}{l}{7 year}&\multicolumn{1}{l}{10 year}&\multicolumn{1}{l}{15 year}&\multicolumn{1}{c}{}&\multicolumn{1}{c}{}&\multicolumn{1}{l}{Joint}&\multicolumn{1}{l}{1 year}&\multicolumn{1}{l}{3 year}&\multicolumn{1}{l}{5 year}&\multicolumn{1}{l}{7 year}&\multicolumn{1}{l}{10 year}&\multicolumn{1}{l}{15 year}\tabularnewline
\midrule
\endhead
%   ~~& \multicolumn{22}{l}{\textbf{TVP NS-VAR}} \tabularnewline
%\midrule
   TVP-NS-VAR &   x&   6&   x&   &   &   &   0.78&   0.91&   0.95&   0.94&   0.89&   0.79&   0.61&   &   &      0.95&   0.99&   1.04&   1.01&   0.98&   0.94&   0.85 \\[-2pt]
   ~~&   &   &   &   &   &   &   (-0.79)&   (-0.28)&   (-0.11)&   (0.03)&   (0.14)&   (0.34)&   (0.72)&   &   &      (-1.42)&   (-0.35)&   (-0.20)&   (-0.06)&   (-0.01)&   (0.01)&   (0.10) \tabularnewline
   ~~&   x&   3&   x&   &   &   &   0.79&   0.96&   1.00&   0.96&   0.90&   0.78&   0.60&   &   &      0.95&   1.03&   1.06&   1.02&   0.98&   0.94&   0.84 \\[-2pt]
   ~~&   &   &   &   &   &   &   (-0.66)&   (-0.21)&   (-0.09)&   (0.04)&   (0.14)&   (0.35)&   (0.73)&   &    &   (-1.35)&   (-0.27)&   (-0.14)&   (-0.03)&   (0.00)&   (0.02)&   (0.11)\tabularnewline
   ~~&   x&   1&   x&   &   &   &   \textbf{0.77}&   0.94&   0.96&   \textbf{0.93}&   \textbf{0.88}&   \textbf{0.77}&   \textbf{0.59}&   &   &     0.95&   1.04&   1.05&   1.01&   0.98&   0.94&   0.84 \\[-2pt]
   ~~&   &   &   &   &   &   &   (-0.67)&   (-0.14)&   (-0.04)&   (0.06)&   (0.15)&   (0.34)&   (0.73)&   &     &   (-1.19)&   (-0.21)&   (-0.10)&   (-0.01)&   (0.01)&   (0.02)&   (0.11)\tabularnewline
\cmidrule{2-23}
   ~~&   x&   6&   --&   &   &   &   0.80&   0.93&   0.98&   0.96&   0.91&   0.79&   0.61&   &     &   0.94&   0.96&   1.00&   0.99&   0.97&   0.94&   0.85 \\[-2pt]
   ~~&   &   &   &   &   &   &   (-0.89)&   (-0.32)&   (-0.16)&   (0.00)&   (0.13)&   (0.34)&   (0.72)&   &     &   (-1.50)&   (-0.39)&   (-0.23)&   (-0.07)&   (-0.01)&   (0.02)&   (0.11)\tabularnewline
   ~~&   x&   3&   --&   &   &   &   0.79&   0.92&   0.98&   0.96&   0.91&   0.79&   0.60&   &     &   0.92&   0.95&   0.99&   0.97&   0.95&   0.92&   0.83 \\[-2pt]
   ~~&   &   &   &   &   &   &   (-0.66)&   (-0.22)&   (-0.10)&   (0.03)&   (0.14)&   (0.35)&   (0.72)&   &     &   (-1.26)&   (-0.27)&   (-0.16)&   (-0.03)&   (0.02)&   (0.04)&   (0.13)\tabularnewline
   ~~&   x&   1&   --&   &   &   &   0.78&   0.92&   0.98&   0.95&   0.89&   0.77&   0.59&   &     &   0.93&   0.96&   0.99&   0.97&   0.96&   0.93&   0.83 \\[-2pt]
   ~~&   &   &   &   &   &   &   (-0.63)&   (-0.18)&   (-0.08)&   (0.05)&   (0.15)&   (\textbf{0.36})&   (0.74)&   &    &   (-1.26)&   (-0.24)&   (-0.13)&   (-0.02)&   (0.02)&   (0.04)&   (0.13)\tabularnewline
\cmidrule{2-23}
   ~~&   --&   6&   x&   &   &   &   0.79&   0.96&   0.97&   0.95&   0.90&   0.79&   0.60&   &     &   0.94&   1.04&   1.03&   1.00&   0.97&   0.93&   0.84 \\[-2pt]
   ~~&   &   &   &   &   &   &   (-0.86)&   (-0.30)&   (-0.11)&   (0.04)&   (0.15)&   (0.35)&   (0.72)&   &     &   (-1.49)&   (-0.37)&   (-0.19)&   (-0.05)&   (0.00)&   (0.01)&   (0.09)\tabularnewline
   ~~&   --&   3&   x&   &   &   &   0.78&   0.95&   \textbf{0.95}&   0.93&   0.89&   0.78&   0.60&     &   &   0.94&   1.02&   1.03&   0.99&   0.96&   0.93&   0.83 \\[-2pt]
   ~~&   &   &   &   &   &   &   (-0.70)&   (-0.21)&   (-0.06)&   (0.06)&   (0.15)&   (0.34)&   (0.72)&   &     &   (-1.25)&   (-0.28)&   (-0.13)&   (-0.02)&   (0.02)&   (0.03)&   (0.11)\tabularnewline
   ~~&   --&   1&   x&   &   &   &   0.78&   0.98&   0.97&   0.94&   0.89&   0.78&   0.60&   &     &   0.92&   0.99&   0.99&   0.97&   0.95&   0.92&   0.83 \\[-2pt]
   ~~&   &   &   &   &   &   &   (-0.70)&   (-0.18)&   (-0.04)&   (0.07)&   (\textbf{0.17})&   (0.36)&   (0.73)&   &   &    (-1.16)&   (-0.22)&   (-0.09)&   (0.00)&   (0.02)&   (0.03)&   (0.12)\tabularnewline
\cmidrule{2-23}
   ~~&   --&   6&   --&   &   &   &   0.79&   0.94&   0.97&   0.95&   0.90&   0.79&   0.61&   &     &   0.93&   0.98&   0.99&   0.97&   0.96&   0.93&   0.84 \\[-2pt]
   ~~&   &   &   &   &   &   &   (-0.91)&   (-0.33)&   (-0.15)&   (0.00)&   (0.12)&   (0.33)&   (0.70)&   &     &   (-1.43)&   (-0.40)&   (-0.23)&   (-0.07)&   (-0.01)&   (0.02)&   (0.10)\tabularnewline
   ~~&   --&   3&   --&   &   &   &   0.79&   0.94&   0.96&   0.95&   0.90&   0.79&   0.60&   &     &   0.92&   0.98&   0.98&   0.96&   0.95&   0.92&   0.83 \\[-2pt]
   ~~&   &   &   &   &   &   &   (-0.86)&   (-0.23)&   (-0.10)&   (0.04)&   (0.15)&   (0.34)&   (0.72)&   &     &   (-1.37)&   (-0.29)&   (-0.15)&   (-0.02)&   (0.02)&   (0.03)&   (0.12)\tabularnewline
   ~~&   --&   1&   --&   &   &   &   0.78&   0.95&   0.97&   0.94&   0.89&   0.78&   0.60&   &   &     \textbf{0.91}&   0.97&   \textbf{0.97}&   0.96&   0.95&   \textbf{0.91}&   \textbf{0.82} \\[-2pt]
   ~~&   &   &   &   &   &   &   (-0.72)&   (-0.20)&   (-0.06)&   (0.05)&   (0.15)&   (0.34)&   (0.72)&   &     &   (-1.30)&   (-0.22)&   (-0.11)&   (0.00)&   (0.03)&   (0.04)&   (0.11)\tabularnewline
\cmidrule{2-23}
      ~~&   x &   -- &   --&   &   &   &   0.78&   0.93&   0.99&   0.95&   0.89&   0.77&   0.59&   &     &   0.91&   \textbf{0.95}&   0.97&   0.96&   \textbf{0.95}&   0.91&   0.82 \\[-2pt]
   ~~&   &   &   &   &   &   &   (-0.36)&   (0.05)&   (0.00)&   (0.07)&   (0.16)&   (0.35)&   (0.74)&   &     &   (-0.95)&   (0.02)&   (-0.02)&   (0.03)&   (\textbf{0.05})&   (0.06)&   (\textbf{0.15})\tabularnewline   
       ~~&  -- &   -- &   x&   &   &   &   0.78&   0.95&   0.98&   0.95&   0.89&   0.78&   0.60&   &   &     0.94&   1.05&   1.03&   0.99&   0.96&   0.93&   0.83 \\[-2pt]
   ~~&   &   &   &   &   &   &   (-0.76)&   (-0.24)&   (-0.07)&   (0.04)&   (0.14)&   (0.33)&   (0.71)&   &     &   (-1.26)&   (-0.29)&   (-0.12)&   (-0.03)&   (0.00)&   (0.01)&   (0.11)\tabularnewline
   ~~&  -- &   $K$ &   --&   &   &   &   0.79&   0.96&   0.98&   0.95&   0.90&   0.78&   0.60&   &     &   0.92&   0.99&   1.00&   0.97&   0.95&   0.92&   0.83 \\[-2pt]
   ~~&   &   &   &   &   &   &   (-0.42)&   (0.11)&   (0.01)&   (0.07)&   (0.15)&   (0.34)&   (0.71)&   &     &   (-1.02)&   (0.07)&   (-0.01)&   (0.02)&   (0.03)&   (0.03)&   (0.11)\tabularnewline
\cmidrule{2-23}
 ~~& \multicolumn{6}{l}{\textbf{Constant}} &   0.79&   1.01&   0.99&   0.96&   0.90&   0.78&   0.60&   &     &   0.91&   0.99&   0.97&   \textbf{0.96}&   0.95&   0.92&   0.82 \\[-2pt]
   ~~&   &   &   &   &   &   &   (-0.82)&   (-0.37)&   (-0.11)&   (0.03)&   (0.13)&   (0.33)&   (0.71)&   &     &   (-1.34)&   (-0.36)&   (-0.13)&   (-0.01)&   (0.02)&   (0.03)&   (0.12)\tabularnewline
   \midrule
    ~~&&&&&&&&&&&&&&&&&&&&&&\\
    ~~&&&&&&&&&&&&&&&&&&&&&&\\
    ~~&&&&&&&&&&&&&&&&&&&&&&\\
    ~~&&&&&&&&&&&&&&&&&&&&&&\\
    ~~&&&&&&&&&&&&&&&&&&&&&&\\
    ~~&&&&&&&&&&&&&&&&&&&&&&\\
    ~~&&&&&&&&&&&&&&&&&&&&&&\\
    ~~&&&&&&&&&&&&&&&&&&&&&&\\
    ~~&&&&&&&&&&&&&&&&&&&&&&\\
    ~~&&&&&&&&&&&&&&&&&&&&&&\\
    ~~&&&&&&&&&&&&&&&&&&&&&&\\
    ~~&&&&&&&&&&&&&&&&&&&&&&\\
    ~~&&&&&&&&&&&&&&&&&&&&&&\\
    ~~&&&&&&&&&&&&&&&&&&&&&&\\
%   ~~& \multicolumn{22}{l}{\textbf{TVP VAR}} \tabularnewline
% \midrule
   TVP-VAR &   x&   6&   x&   &   &   &   0.79&   0.92&   0.96&   0.94&   0.89&   0.79&   0.60&   &   &     0.94&   0.96&   0.99&   0.98&   0.97&   0.94&   0.84 \\[-2pt]
   ~~&   &   &   &   &   &   &   (1.13)&   (0.06)&   (0.03)&   (0.07)&   (0.14)&   (0.33)&   (0.68)&   &     &   (0.63)&   (-0.01)&   (-0.01)&   (0.00)&   (0.00)&   (0.01)&   (0.08)\tabularnewline
   ~~&   x&   3&   x&   &   &   &   0.78&   0.92&   0.96&   0.94&   0.89&   0.78&   0.60&   &   &     0.94&   0.98&   1.00&   0.99&   0.98&   0.94&   0.84 \\[-2pt]
   ~~&   &   &   &   &   &   &   (1.20)&   (0.09)&   (0.03)&   (0.06)&   (0.14)&   (0.33)&   (0.70)&   &     &   (0.55)&   (0.01)&   (-0.02)&   (-0.01)&   (0.00)&   (0.02)&   (0.10)\tabularnewline
   ~~&   x&   1&   x&   &   &   &   0.79&   0.93&   0.97&   0.95&   0.89&   0.78&   0.60&   &   &     0.93&   0.96&   0.99&   0.98&   0.97&   0.94&   0.84 \\[-2pt]
   ~~&   &   &   &   &   &   &   (1.52)&   (0.10)&   (0.03)&   (0.06)&   (0.14)&   (0.33)&   (0.70)&   &      &   (\textbf{0.83})&   (0.02)&   (0.00)&   (0.01)&   (0.01)&   (0.02)&   (0.12)\tabularnewline
\cmidrule{2-23}
%~~&&&&&&&&&&&&&&&&&&&&&&&\\
   ~~&   x&   6&   --&   &   &   &   0.79&   0.93&   0.98&   0.95&   0.90&   0.79&   0.60&   &   &     0.94&   0.99&   1.00&   1.00&   0.98&   0.95&   0.84 \\[-2pt]
   ~~&   &   &   &   &   &   &   (1.26)&   (0.06)&   (0.04)&   (0.07)&   (0.15)&   (0.34)&   (0.69)&   &     &   (0.69)&   (0.01)&   (0.01)&   (0.02)&   (0.02)&   (0.03)&   (0.10)\tabularnewline
   ~~&   x&   3&   --&   &   &   &   0.78&   0.91&   0.96&   0.94&   0.89&   0.79&   0.60&   &   &     0.94&   0.98&   1.00&   1.00&   0.98&   0.95&   0.85 \\[-2pt]
   ~~&   &   &   &   &   &   &   (1.20)&   (0.11)&   (\textbf{0.06})&   (\textbf{0.09})&   (0.16)&   (0.34)&   (0.71)&   &   &     (0.66)&   (0.03)&   (0.02)&   (0.02)&   (0.03)&   (0.04)&   (0.12)\tabularnewline
   ~~&   x&   1&   --&   &   &   &   0.78&   \textbf{0.91}&   0.96&   0.94&   0.89&   0.78&   0.60&   &     &   0.92&   0.96&   0.97&   0.97&   0.96&   0.93&   0.83 \\[-2pt]
   ~~&   &   &   &   &   &   &   (1.56)&   (0.13)&   (0.06)&   (0.08)&   (0.15)&   (0.34)&   (0.72)&   &     &   (0.61)&   (0.07)&   (\textbf{0.05})&   (0.04)&   (0.04)&   (0.05)&   (0.14)\tabularnewline
\cmidrule{2-23}
   ~~&   --&   6&   x&   &   &   &   0.81&   0.95&   0.97&   0.95&   0.91&   0.81&   0.63&   &   &      0.97&   1.00&   1.00&   1.00&   1.00&   0.98&   0.90 \\[-2pt]
   ~~&   &   &   &   &   &   &   (0.95)&   (0.02)&   (0.03)&   (0.07)&   (0.13)&   (0.32)&   (0.67)&   &      &   (0.45)&   (-0.04)&   (-0.03)&   (-0.01)&   (-0.01)&   (0.00)&   (0.06)\tabularnewline
   ~~&   --&   3&   x&   &   &   &   0.79&   0.95&   0.98&   0.95&   0.90&   0.79&   0.60&   &   &      0.93&   0.99&   0.99&   0.97&   0.96&   0.93&   0.83 \\[-2pt]
   ~~&   &   &   &   &   &   &   (1.08)&   (0.06)&   (0.02)&   (0.06)&   (0.12)&   (0.31)&   (0.68)&   &      &   (0.28)&   (-0.01)&   (-0.02)&   (-0.01)&   (-0.01)&   (0.01)&   (0.08)\tabularnewline
   ~~&   --&   1&   x&   &   &   &   0.78&   0.95&   0.96&   0.94&   0.89&   0.79&   0.60&   &   &      0.95&   0.97&   0.98&   0.98&   0.98&   0.96&   0.87 \\[-2pt]
   ~~&   &   &   &   &   &   &   (1.29)&   (0.08)&   (0.04)&   (0.07)&   (0.14)&   (0.32)&   (0.69)&   &      &   (0.44)&   (0.04)&   (0.00)&   (-0.01)&   (-0.01)&   (0.00)&   (0.09)\tabularnewline
\cmidrule{2-23}
   ~~&   --&   6&   --&   &   &   &   0.79&   0.95&   0.96&   0.95&   0.90&   0.79&   0.60&   &      &   0.92&   0.99&   0.98&   0.97&   0.96&   0.93&   0.83 \\[-2pt]
   ~~&   &   &   &   &   &   &   (1.04)&   (0.04)&   (0.05)&   (0.08)&   (0.15)&   (0.34)&   (0.69)&      &   &   (0.32)&   (-0.03)&   (0.01)&   (0.02)&   (0.03)&   (0.04)&   (0.11)\tabularnewline
   ~~&   --&   3&   --&   &   &   &   0.79&   0.94&   0.97&   0.94&   0.89&   0.79&   0.61&   &     &   0.93&   0.98&   0.98&   0.97&   0.96&   0.93&   0.83 \\[-2pt]
   ~~&   &   &   &   &   &   &   (1.27)&   (0.08)&   (0.05)&   (0.08)&   (0.15)&   (0.34)&   (0.70)&   &      &   (0.61)&   (0.00)&   (0.03)&   (0.04)&   (0.05)&   (0.06)&   (0.13)\tabularnewline
   ~~&   --&   1&   --&   &   &   &   0.79&   0.95&   0.96&   0.95&   0.90&   0.79&   0.61&   &      &   0.92&   0.99&   0.99&   0.97&   0.95&   0.92&   0.82 \\[-2pt]
   ~~&   &   &   &   &   &   &   (\textbf{1.58})&   (0.09)&   (0.06)&   (0.08)&   (0.15)&   (0.35)&   (0.72)&      &   &   (0.78)&   (0.04)&   (0.04)&   (\textbf{0.05})&   (0.05)&   (\textbf{0.06})&   (0.15)\tabularnewline
\cmidrule{2-23}
   ~~&   x &   -- &   --&   &   &   &     0.79&   0.93&   0.97&   0.95&   0.90&   0.79&   0.60&   &   &      0.93&   0.98&   0.99&   0.98&   0.97&   0.93&   0.83 \\[-2pt]
   ~~&   &   &   &   &   &   &   (1.26)&   (0.12)&   (0.04)&   (0.07)&   (0.14)&   (0.35)&   (\textbf{0.75})&      &   &   (0.75)&   (0.07)&   (0.03)&   (0.02)&   (0.01)&   (0.03)&   (0.15)\tabularnewline
   ~~&  -- &   -- &   x &   &   &   &     0.79&   0.94&   0.96&   0.93&   0.89&   0.79&   0.61&   &   &      0.93&   0.97&   0.97&   0.97&   0.96&   0.94&   0.86 \\[-2pt]
   ~~&   &   &   &   &   &   &   (0.83)&   (0.07)&   (0.02)&   (0.07)&   (0.14)&   (0.33)&   (0.70)&   &      &   (0.49)&   (0.02)&   (0.00)&   (0.01)&   (0.01)&   (0.02)&   (0.11)\tabularnewline
   ~~&  -- &  $K$ &   --&   &   &   &     0.79&   0.94&   0.96&   0.95&   0.90&   0.80&   0.61&   &      &   0.93&   0.98&   0.98&   0.97&   0.96&   0.93&   0.83 \\[-2pt]
   ~~&   &   &   &   &   &   &   (0.28)&   (\textbf{0.17})&   (0.03)&   (0.04)&   (0.10)&   (0.27)&   (0.67)&      &   &   (-0.15)&   (\textbf{0.12})&   (0.01)&   (-0.02)&   (-0.03)&   (-0.04)&   (0.07)\tabularnewline
\cmidrule{2-23}
\shadeBench   ~~& \multicolumn{6}{l}{\textbf{Constant}} &   0.85&   0.39&   0.59&   0.72&   0.82&   0.99&   1.28&   &   &      0.76&   0.43&   0.62&   0.73&   0.81&   0.87&   0.96 \\[-2pt]
\shadeBench   ~~&   &   &   &   &   &   &   (2.04)&   (-0.21)&   (-0.81)&   (-1.08)&   (-1.25)&   (-1.49)&   (-1.87)&   &   &      (2.26)&   (-0.30)&   (-0.84)&   (-1.08)&   (-1.19)&   (-1.26)&   (-1.35)\tabularnewline
\bottomrule
\caption*{\scriptsize\textit{Notes}: We present the results of out-of-sample one-month-ahead and one-quarter-ahead forecasting from the 15 TVP-NS-VAR and 15 TVP-VAR model variants for maturities 1, 3, 5, 7, 10 and 15 years and the corresponding joint measure across all maturities. Specifications of the TVP model variants are differentiated over a grid of effect modifier combinations, in terms of whether the exogenous variables in $\bm{r}_t$ are included or not (``x'' marks inclusion, ``--'' indicates absence), the number $\delta= R_{\tau}/M$ of latent factors per equation (with $\delta = K$ yielding conventional independent random walk specifications for the TVPs) and the presence of Markov switching factors in $\bm{S}_t$, again with ``x'' marking inclusion and ``--'' absence. ``Constant'' labels a conventional constant parameter VAR. $M=3$ and $K=9$ in the case of NS-VAR, and $M=6$ and $K=18$ in the case of VAR. We estimate and forecast recursively, using data from 1973:01 to the time that the forecast is made, beginning in 2000:01 through 2019:12. Root Mean Squared Errors (RMSEs) and Log Predictive Bayes Factors (LPBFs), averaged over the out-of-sample observations are given relative to the constant parameter VAR model (shaded in yellow). The best performing model specification by column is given in bold, highlighting the specification with the smallest RMSE ratio and the largest positive LPBF difference, respectively.}
\end{longtable}

\end{tiny}
\end{landscape}
\clearpage%
}

The performance of point forecasts is measured in terms of RMSEs and that of density forecasts in terms of LPBFs, relative to the constant parameter VAR model (shaded in yellow) that serves as benchmark. RMSEs are presented in ratios, and LPBFs in differences are given below the RMSEs in parentheses. RMSEs below one indicate superior performance, relative to the benchmark, as LPBF figures greater than zero do. The best performing model specification by column is given in bold, highlighting the specification with the smallest RMSE ratio and the largest positive LPBF difference, respectively.

The vast range of competing specifications, loss functions used to evaluate forecasts and maturities makes it hard to identify a single best performing model. We first provide a general overview on model performance and then zoom into differences in predictive accuracy for point and density forecasts.

At a very general level, Table \ref{tab:LPS} suggests a pronounced degree of heterogeneity in forecast accuracy across models and for both the NS-VAR and VAR specifications. While differences at some maturities are substantial, they are muted or non-existent for others. It is also worth mentioning that the TVP variants of the NS-VAR and VAR models outperform the constant parameter specifications in most cases (apart from one-quarter ahead point forecasts of treasuries with a maturity of five years). This general observation holds true irrespective of whether only point forecasts or the full predictive distribution are considered. These accuracy premia point towards the necessity  of  addressing structural breaks in the dynamic evolution of the yield curve.

Comparing the NS-VAR and VAR specifications indicates that the latter usually perform better for density forecasts, while the former are often superior in terms of point forecasts. The better point forecasting performance of the NS-VAR suggests that the three factors contain relevant information for the first moment of the predictive distribution. When we also consider higher order moments, this story changes. The better density performance of the VAR models is most likely driven by two sources. The first is that, as opposed to the conditional mean, the strong implicit assumptions of the NS-VAR on the prediction error variance-covariance matrix seems overly restrictive. Allowing for richer dynamics in the covariances by explicitly modeling the shocks to a panel of yields thus yields better density forecasts. The second fact is that the small-scale NS-VARs might feature a too tight predictive variance since relevant information is ignored. And this could harm density forecasting accuracy during turbulent times.

Next, we zoom into specific model classes. Within these,  differences in performance along sub-divisions (provided by the inclusion/exclusion of the exogenous variables or the Markov switching processes and the number of latent random walk factors) are often negligible. This finding indicates that our flexible approach to probabilistically selecting the most adequate state evolution via Bayesian shrinkage is successful, and thus not susceptible to overfitting concerns.

We now turn to considering model-specific forecasting accuracy with the aim to find the best performing models for point and density forecasts and for the two different forecast horizons. The overall winner for point forecasts at the one-month horizon, on average, is the flexible NS-VAR model specification featuring the exogenous variables, one latent factor per equation and the Markov switching processes. While this specification yields RMSEs that are $23$ percent lower than those of the benchmark specification, it must be acknowledged that most competing specifications exhibit values that are similar in magnitude. The same is true for the best-performing model specification at the one-quarter-ahead horizon, the one-factor NS-VAR without exogenous variables and Markov switching processes, albeit at much smaller margins versus the benchmark. Here, improvements are about nine percent in terms of relative RMSEs. Assessing predictive performance for individual maturities allows to identify which segment of the yield curve drives the overall results. While gains for shorter maturities in the case of one-month-ahead forecasts are muted, we observe large gains at the long end of the yield curve. Relative to the best performing specification, improvements are about $40$ percent in RMSE terms for the best performing specification. The same is true for one-quarter-ahead forecasts, however, at smaller margins of about $20$ percent.

We proceed with our findings for density forecasts. As mentioned above, the VAR specifications overall exhibit more favorable relative LPBFs compared to the NS-VAR. This is easily observable by noting that most bold values (in parentheses) are located in the lower panel of Table \ref{tab:LPS}. In terms of average performance at the one-month horizon, we find that the TVP-VAR specification with one factor but no exogenous variables or Markov switching performs best, closely followed by the most flexible specification with one unobserved factor including Markov switching. This again serves as an example that even though it is extremely flexible, our approach to shrinking the parameter space avoids overfitting and does not harm predictive performance. In fact, for one-quarter-ahead forecasts, the TVP-VAR with exogeneous variables, one latent factor and Markov switching shows the largest gains in terms of density forecasts. Again, it must be acknowledged that margins within this model class are rather small. As it is the case for point forecasts, these gains are mostly obtained in terms of the long end of the yield curve, while gains in forecast accuracy at shorter horizons are negligible.

Summing up, while improvements relative to established models are often small, our proposed framework is competitive for most maturities at both the one-month and one-quarter ahead horizons. Main differences arise from the considered model class, with the NS-VAR exhibiting promising results in terms of point forecasts, and superior density forecasts for the VAR. Moreover, we detect the largest improvements at the long end of the yield curve. The additional flexibility of our proposed approach does not lead to severe overfitting since we use shrinkage priors to regularize several parts of the parameter space. And it almost never harms forecast accuracy while improving performance in some cases. 

% \onehalfspacing
% \begin{landscape}
% \begin{tiny}
% \input{tabs/rmse_lps_new_altern.tex}
% \end{tiny}
% \end{landscape}

\doublespacing

\subsection{Determinants of Time-Variation in the Coefficients}\label{sec:insample}

The previous sub-section  established that the proposed approach yields more favorable predictions than  conventional TVP-VARs. Including observed and latent effect modifiers allows for investigating the sources of time-variation in coefficients, and thus the driving factors of improvements in predictive accuracy. We carry out a detailed analysis of these determinants in  this sub-section.

Because of its favorable forecasting properties, we choose the NS-VAR with one latent factor per equation to investigate the driving forces of parameter variation over the full estimation sample. Recall that the choice of $\bm z_t$ for this specification translates into a single equation-specific latent factor ($\tau_{jt}$), an equation-specific Markov switching indicator $S_{jt}$ and the three observed early warning indicators. %, including the risk-free interest rate (\texttt{RF}), the \texttt{NFCI} index and the NBER recession indicator (labeled \texttt{REC}).

% Fig 1: Factors
To illustrate the observed and latent factors, we transform each column vector in $\bm z_t$ ex-post such that it is bounded between zero and one. This allows for comparisons even though some blocks of the respective matrices are not econometrically identified.\footnote{In particular, this is the case for the product $\bm \Lambda_\tau \bm \tau_t$. Note that we do not face this issue for $\bm \Lambda_r \bm r_t$ and $\bm \Lambda_S \bm S_t$ because both are either observed or already bounded between zero and one, and their scale and sign are identified.} We therefore exploit the fact that introducing any invertible $R \times R$-dimensional matrix does not alter the likelihood of the model, since $\bm \Lambda \bm U^{-1} \bm U \bm z_t  = \bm \Lambda \bm z_t$. Define $\bm U$ as diagonal matrix such that the maximum of each $\bm z_j$ (for $j = 1, \dots, R$)  corresponds to one, the minimum is zero, $\tilde{\bm \Lambda} = \bm \Lambda \bm U^{-1}$ and $\tilde{\bm z}_t = \bm U \bm z_t$. This simple linear transformation allows for assessing the relative movement of the indicators in $\tilde{\bm z}_t$ without affecting overall dynamics. 

\begin{figure}[!htbp]
\centering
\caption{Evolution of the normalized effect modifiers $\tilde{\bm{z}}_t=\bm{U}\bm{z}_t$ over time.\label{fig:factors}}
\begin{minipage}{\textwidth}
\centering
(a) \textit{Observed factors}
\end{minipage}
\begin{minipage}{\textwidth}
\centering
\hspace{2mm}%
\includegraphics[scale=0.43]{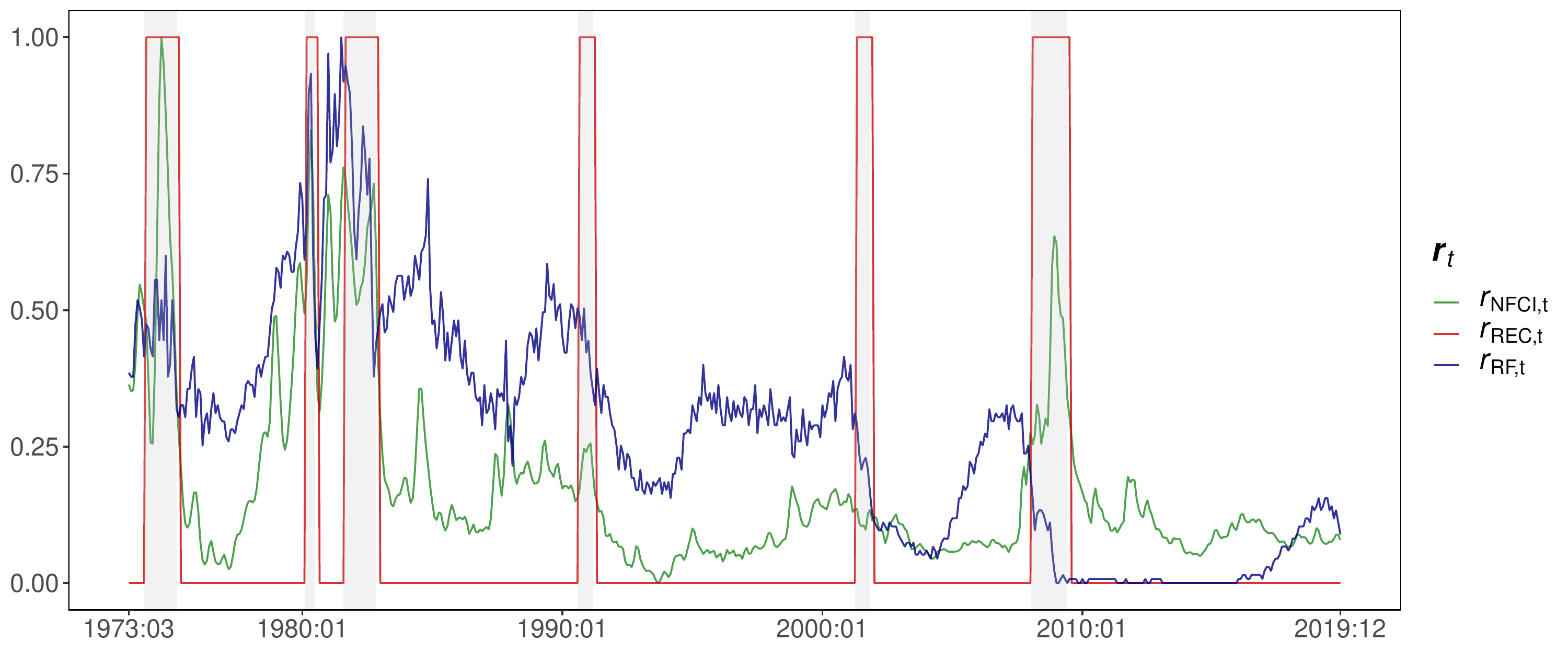}
\end{minipage}
\begin{minipage}{\textwidth}
\centering
\vspace{5pt}
(b) \textit{Equation-specific latent Markov switching factors}
\end{minipage}
\begin{minipage}{\textwidth}
\centering
\includegraphics[scale=0.42]{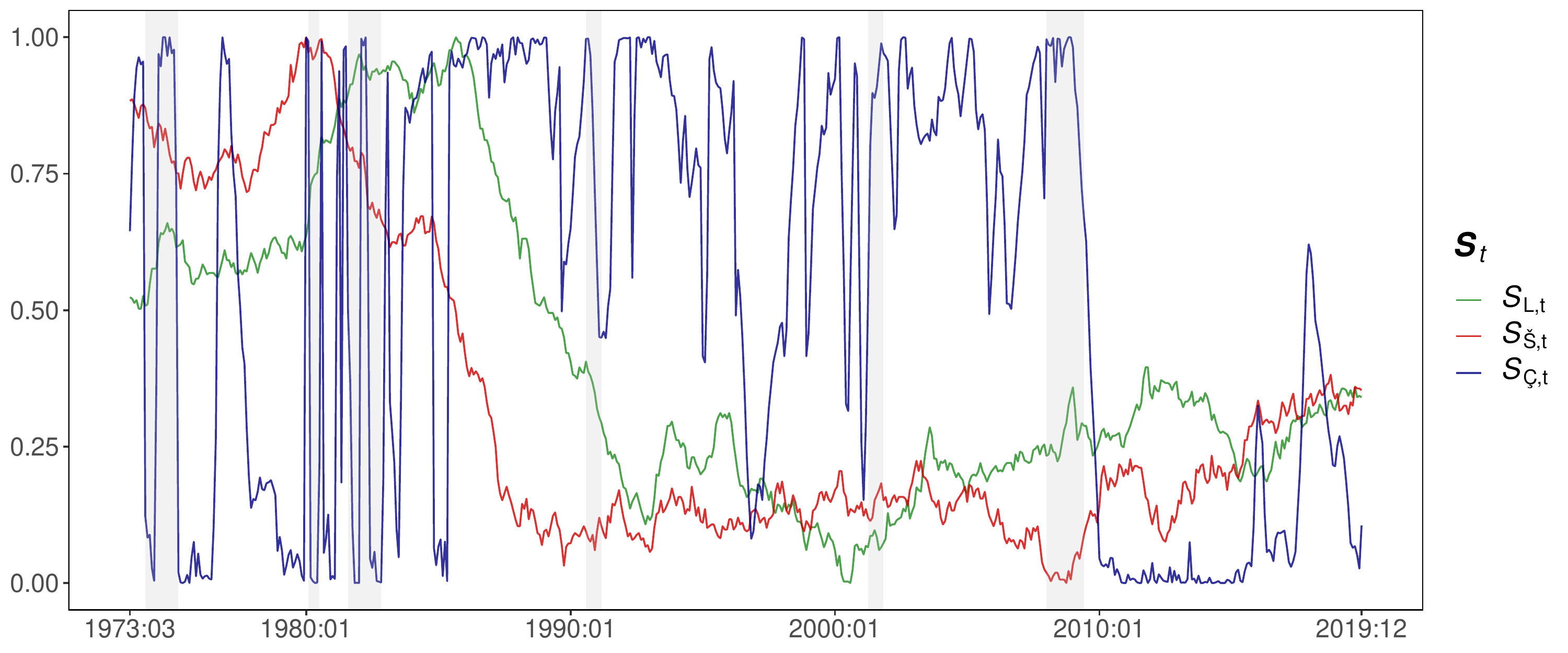}
\end{minipage}
\begin{minipage}{\textwidth}
\centering
\vspace{5pt}
(c) \textit{Equation-specific latent random walk factors}
\end{minipage}
\begin{minipage}{\textwidth}
\centering
\includegraphics[scale=0.415]{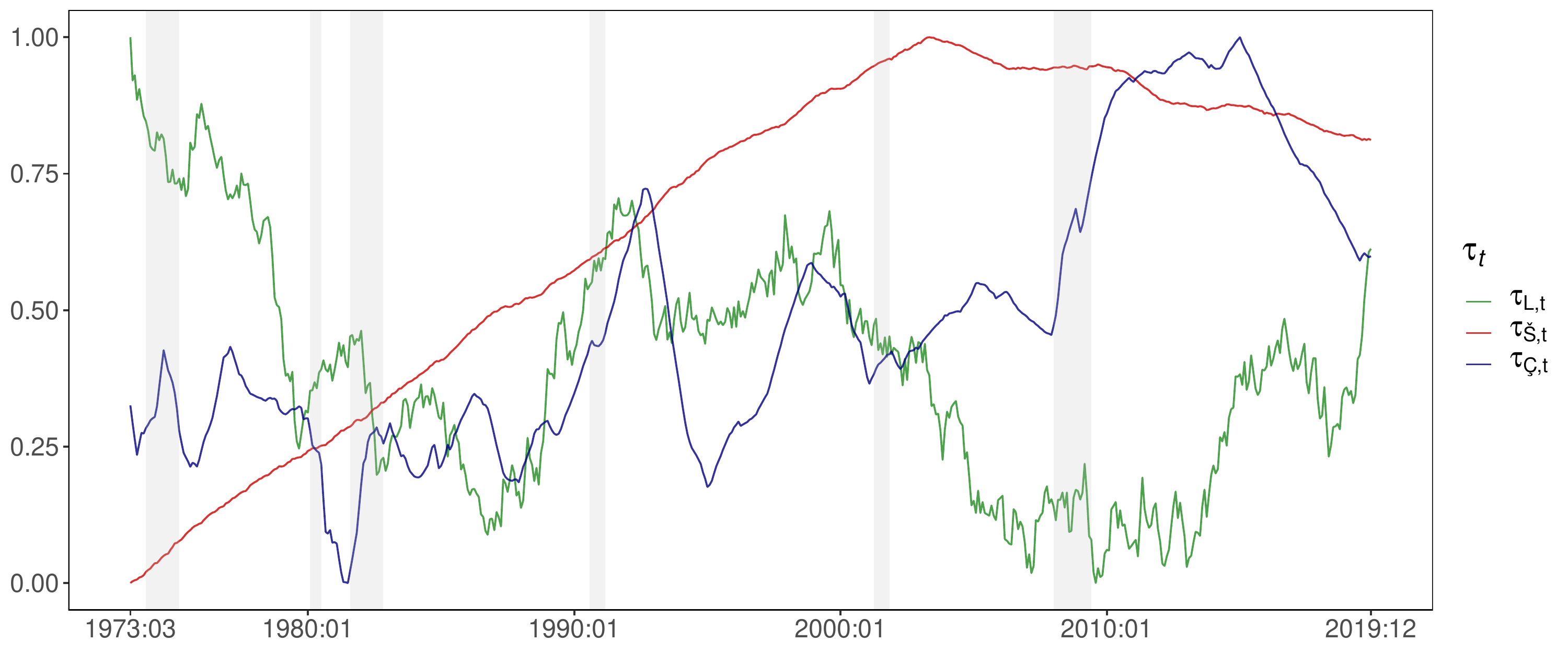}
\end{minipage}
\begin{minipage}{\textwidth}
\end{minipage}
\caption*{\footnotesize \noindent \textit{Notes}: Results are based on a TVP-NS-VAR model specification $\bm{z}_t=(\bm{r}^{\prime}_t,\bm{S}_t^\prime, \bm{\tau}_t^\prime)^\prime$ and $\delta=R_\tau / M=1$ and $P=3$. Panel (a) shows the normalized observed factors $r_{\text{NFCI},t}, r_{\text{REC},t}$ and $r_{\text{RF},t}$ (collected in $\bm{r}_{t}$), panel (b) the posterior mean of the latent Markov switching factors, $S_{L,t}, S_{\text{\textit{\v{S}}},t}$ and $S_{\text{\textit{\c{C}}},t}$ (collected in $\bm{S}_t$) and panel (c) the posterior median of the latent random walk factors, $\tau_L, \tau_\text{\textit{\v{S}}}$ and $\tau_\text{\textit{\c{c}}}$ (collected in $\bm{\tau}_t$). Note that $S_{jt}$ and $\tau_{jt}$ are equation-specific latent quantities with $M(=3)$ endogenous variables, $j \in \{L,\text{\textit{\v{S}}}, \text{\textit{\c{C}}}\}$. $S_{L,t}$ and $\tau_{L,t}$ correspond to the first equation, $S_{\text{\textit{\v{S}}},t}$ and $\tau_{\text{\textit{\v{S}}},t}$ to the second, while $S_{\text{\textit{\c{C}}},t}$ and $\tau_{\text{\textit{\c{C}}},t}$ to the third. Results are based on the TVP-NS-VAR model variant with $\delta =R_\tau / M=1$ and using 15,000 MCMC draws. The gray shaded vertical bars represent recessions dated by the NBER Business Cycle Dating Committee. Sample period: 1973:01 to 2019:12. Vertical axis: normalized values. Front axis: months.}
\end{figure}

Figure \ref{fig:factors} displays the evolution of the normalized indicators in $\tilde{\bm z}_t$ over time. First, we focus on the features of the observed effect modifiers depicted in the upper panel (a). The normalized NFCI index peaks during the oil crisis in the mid $1970$s, while exhibiting a stable evolution at a quite low level during the Great Moderation (the period from $1990$ until to the onset of the global financial crisis in $2007$). Spikes in $r_{\text{NFCI}, t}$ tend to coincide with recessionary episodes, indicated by the binary recession indicator $r_{\text{REC},t}$. The risk-free interest rate $r_{\text{RF},t}$ can be related to the monetary policy stance. Early in the sample we observe substantial increases, peaking during the Volcker disinflation in the early $1980$s. Subsequently, large and abrupt decreases are notable during recessions, while an overall decreasing trend is observable. Before turning to the latent indicators in $\bm S_t$ and $\bm \tau_t$, note that the respective elements $S_{jt}$ and $\tau_{jt}$ are equation-specific with $j\in\{L, \text{\textit{\v{S}}},\text{\textit{\c{C}}}\}$ referring to the level, slope and curvature of the yield curve. The middle panel (b) indicates the posterior median of the three Markov switching factors collected in $\bm S_t$. The lower panel (c) shows the posterior medians of the three (transformed) gradually changing latent random walk factors in $\bm \tau_t$. 

Several features of the latent indicators are worth highlighting. Each of the latent quantities exhibits distinct dynamics and thus carries information in addition to the observed indicators. Examining the posterior means, which are essentially the unconditional posterior probabilities that a given Markov indicator equals one, we observe that both $S_{L,t}$ and $S_{\text{\textit{\v{S}}},t}$ evolve comparatively gradually  over time. Before the Volcker disinflation, both tend to increase with posterior medians above $0.5$, indicating that regime $1$ is more likely. After $1985$, we observe a major shift towards regime $0$. Interestingly, for $S_{\text{\textit{\v{S}}},t}$ this transition appears immediately after $1985$, while we detect a notable delay in $S_{L,t}$.

During the Great Moderation both indicators tend to remain associated with regime $0$. The indicator associated with the middle segment of the yield curve, $S_{\text{\textit{\c{C}}},t}$, by contrast, transitions between regimes at a higher frequency. Particularly during the Volcker disinflation we observe mixed patterns and no clear or steady tendency towards a single regime. This changes between $1990$ and $1997$, where the posterior mean of $S_{\text{\textit{\c{C}}},t}$ is consistently above $0.5$, albeit with several high-frequency movements. In the aftermath of the global financial crisis, with short-term rates approaching the zero lower bond, $S_{\text{\textit{\c{C}}},t}$ switches abruptly into regime $0$. Conditional on the respective loadings in $\bm \Lambda_S$ being non-zero, this feature would directly relate to the observed narrowing spread of the yield curve and accompanying structural breaks in coefficients of the equation related to the curvature of the yield curve. 

We observe several interesting features of the unobserved factors in $\bm{\tau}_t$. While $\tau_{L,t}$ is noisy and indicates substantial high-frequency movements, $\tau_{\text{\textit{\v{S}}},t}$ and $\tau_{\text{\textit{\c{C}}},t}$ are much smoother. The factor governing coefficients in the level-equation of the yield curve peaks early in the sample, followed by a decline between 1980 and 1990. After a brief increase and stabilization between 1990 and 2000, we see gradual declines until the global financial crisis starting in 2007. Since then, the factor shows upward trending movement, with several high-frequency troughs. By contrast, the unobserved factor related to $\mathcal{S}_t$ exhibits approximately linear trending behavior from the beginning of the sample until the early 2000s, where it plateaued. After 2010, a gradual but moderate decrease is visible. $\tau_{\text{\textit{\c{C}}},t}$ is comparable to $\tau_{L,t}$, albeit with several differences. While several peaks coincide, we also find adverse movements, for instance in the brief early 1980s recession and after 2000. Interestingly, high-frequency movements are muted when compared to $\tau_{L,t}$.

\begin{figure}[!hbt]
\centering
\caption{Heat maps for rescaled loadings in $\tilde{\bm{\Lambda}}=
\bm{\Lambda U}^{-1}$.\label{fig:loadings}}
\begin{minipage}{\textwidth}
\centering
(a) \textit{Coefficients}
\vspace{5pt}
\end{minipage}
\begin{minipage}{0.3\textwidth}
\centering
\textit{${L}_t$-equation}
\end{minipage}
\begin{minipage}{0.3\textwidth}
\centering
\textit{$\text{\textit{\v{S}}}_t$-equation}
\end{minipage}
\begin{minipage}{0.3\textwidth}
\centering
\textit{$\text{\textit{\c{C}}}_t$-equation}
\end{minipage}
\begin{minipage}{0.05\textwidth}
\centering
\textcolor{white}{Space keeper}
\end{minipage}
\begin{minipage}{0.3\textwidth}
\centering
\includegraphics[scale=0.45]{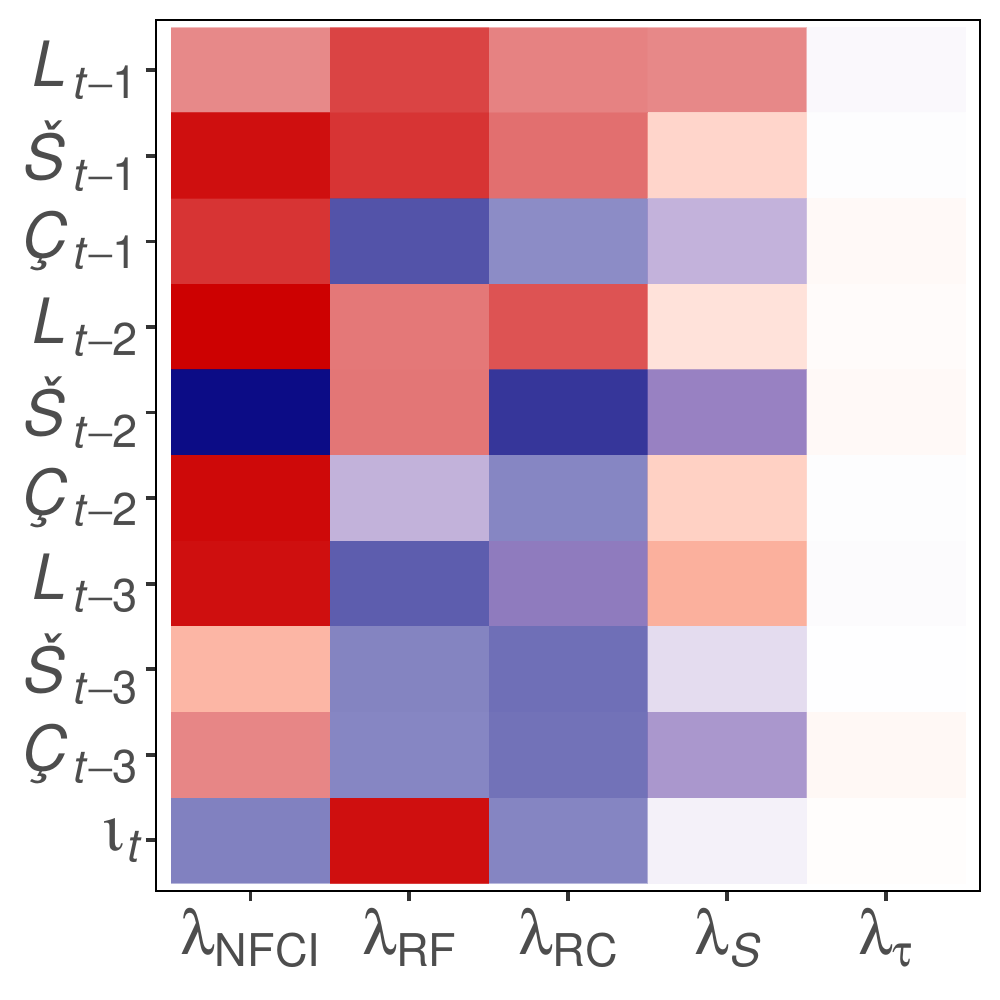}
\end{minipage}
\begin{minipage}{0.3\textwidth}
\centering
\includegraphics[scale=0.45]{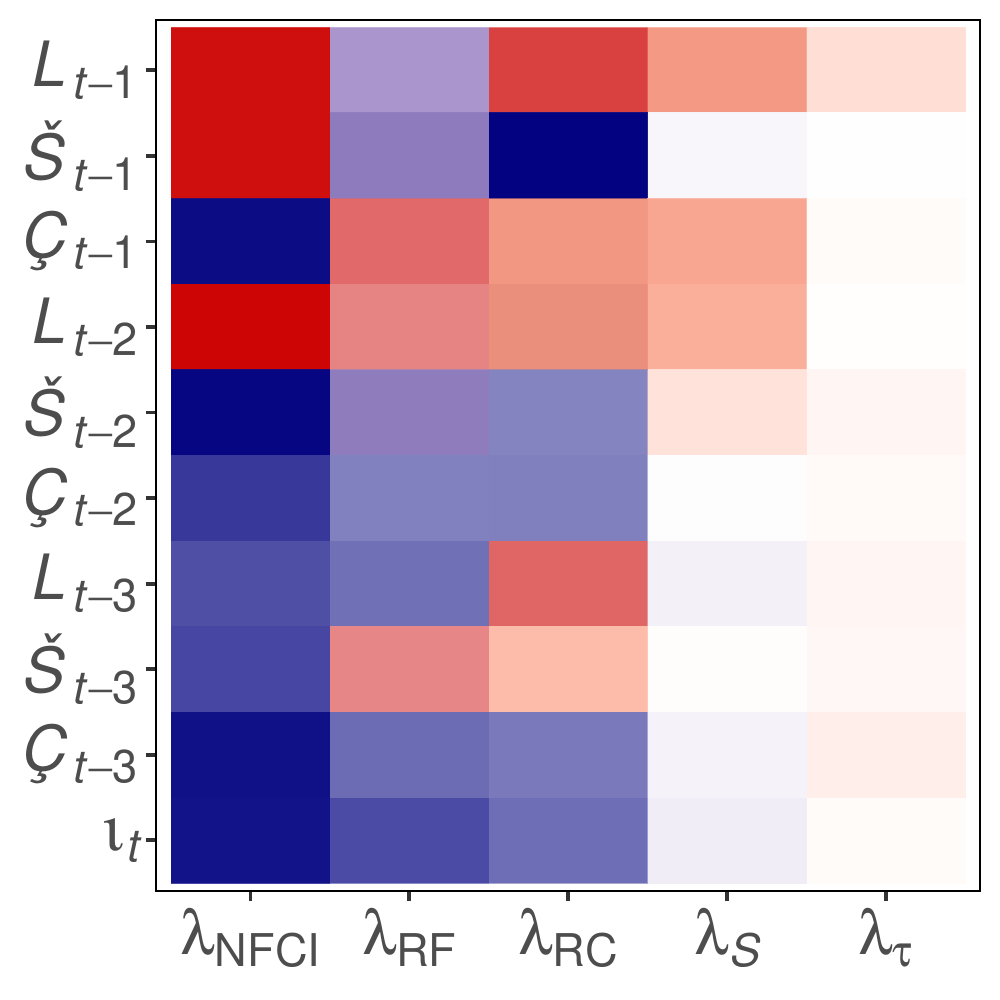}
\end{minipage}
\begin{minipage}{0.3\textwidth}
\centering
\includegraphics[scale=0.45]{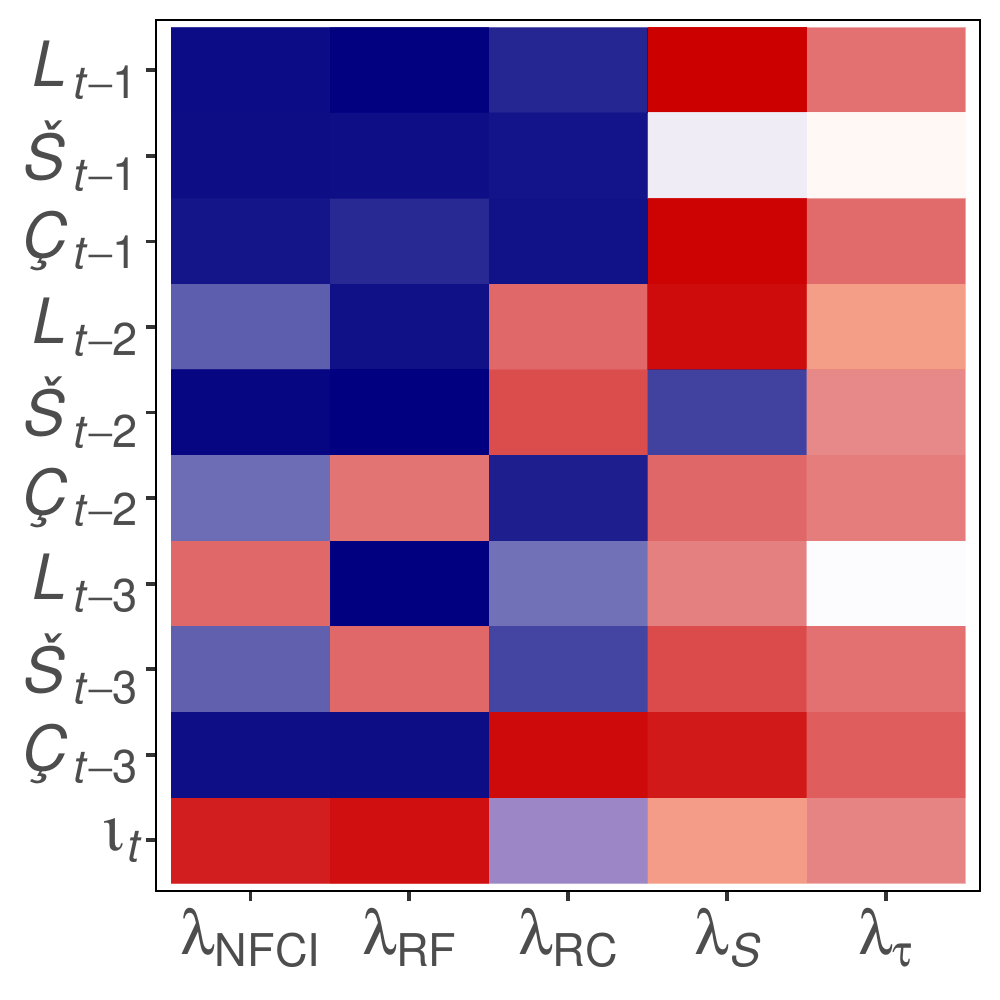}
\end{minipage}
\begin{minipage}{0.05\textwidth}
\hspace*{1pt}%
\includegraphics[scale=0.55]{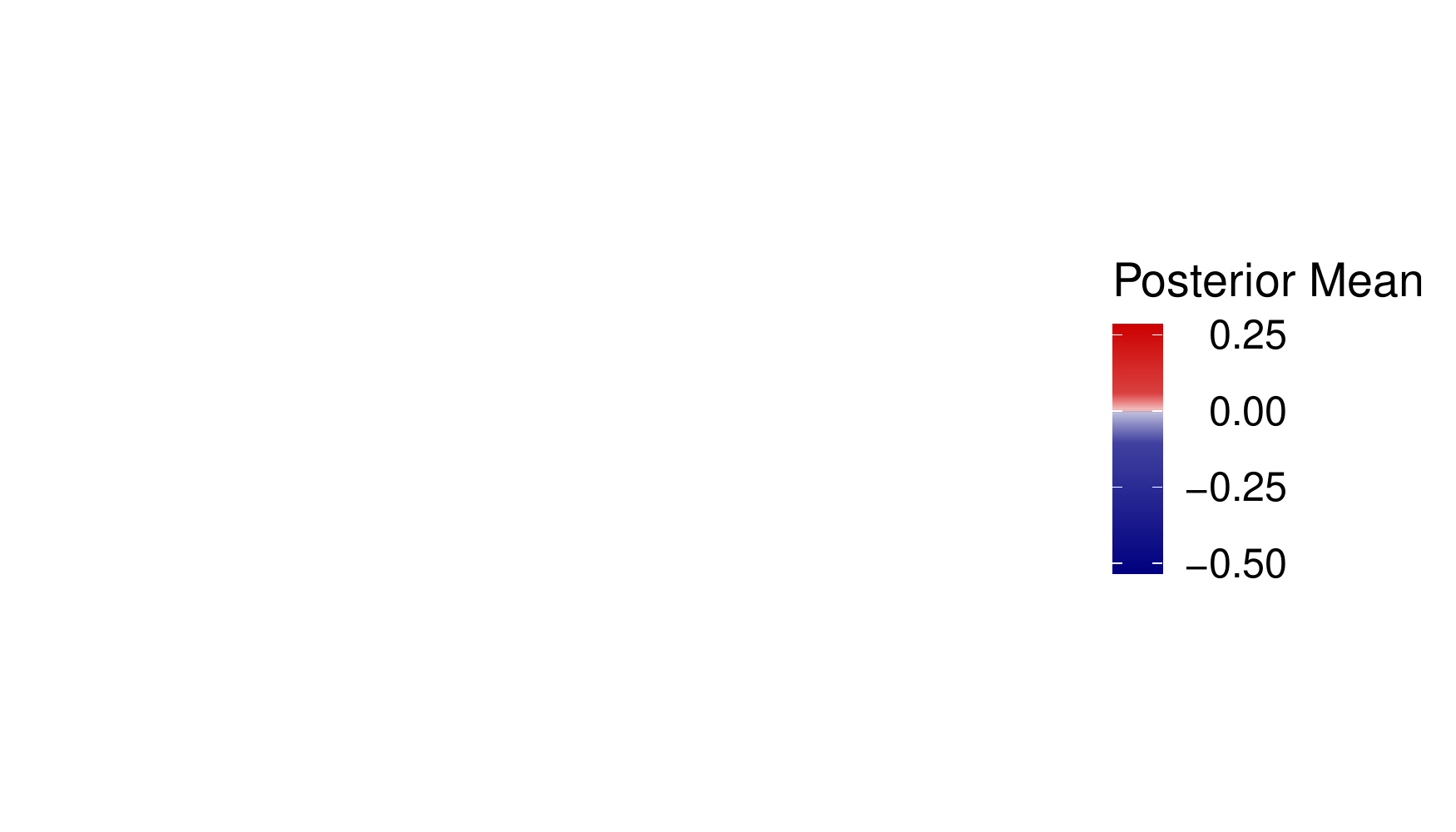}
\end{minipage}
\begin{minipage}{\textwidth}
\centering
\vspace{10pt}
(b) \textit{Covariances}
\vspace{5pt}
\end{minipage}
\begin{minipage}{\textwidth}
\centering
\hspace*{25pt}%
\includegraphics[scale=0.55]{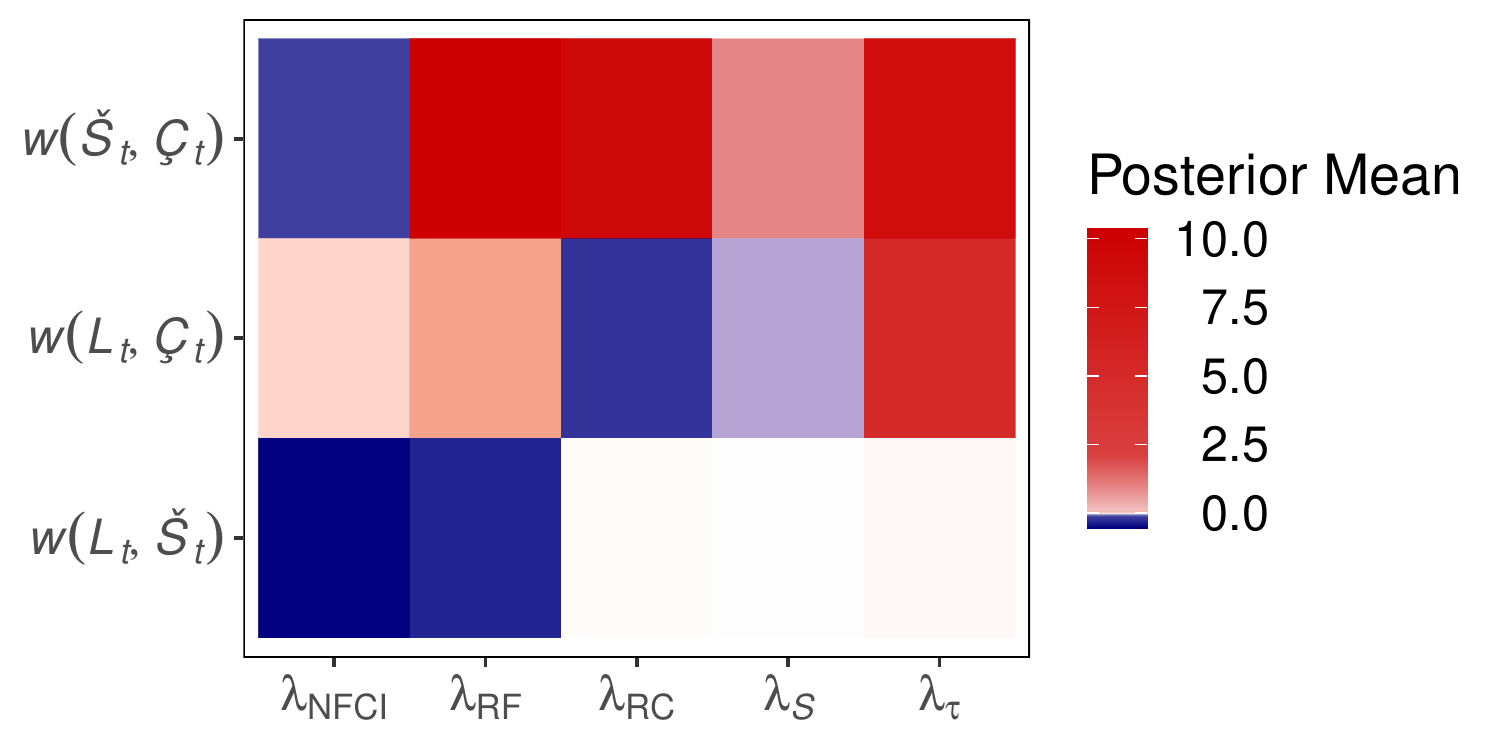}
\end{minipage}
\begin{minipage}{\textwidth}
\end{minipage}
\caption*{\footnotesize\textit{Notes}: $\tilde{\bm{\Lambda}}$ translates the law of motion captured in $\tilde{\bm{z}}_t$ to (a) the VAR coefficients in $\bm{\beta}_t$, and (b) the covariances stored in $\bm{q}_t$, based on a TVP-NS-VAR model specification with $\bm{z}_t=(\bm{r}^{\prime}_t,\bm{S}_t^\prime, \bm{\tau}_t^\prime)^\prime$, $\delta=R_\tau/M=1$ and $P=3$. $\iota_t$ denotes the time-varying intercept, and $L_{t-p},\text{\textit{\v{S}}}_{t-p},\text{\textit{\c{C}}}_{t-p}$ for $p=1,2,3$ the lagged variables $L_t, \text{\textit{\v{S}}}_t$ and $\text{\textit{\c{C}}}_t$, respectively. $\bm{\lambda}_r$ relates to elements of $\bm{\tilde{\Lambda}}$ associated with $r_{\text{NFCI},t}, r_{\text{REC},t}$ and $r_{\text{RF},t}$ and the respective equations, $\bm {\lambda}_{S}$ denotes loadings related to a single Markov switching factor collected in $\bm S_t$, and $\bm\lambda_\tau$ loadings corresponding to latent random walk factors in $\bm\tau_t$. Note that $S_{jt}$ and $\tau_{jt}$ for $j \in \{L,\text{\textit{\v{S}}},\text{\textit{\c{C}}}\}$ are equation-specific quantities, while $\bm{r}_t$ stays fixed across equations. $w(\text{\textit{\v{S}}}_t, \text{\textit{\c{C}}}_t)$ denotes the contemporaneous relation between the $\text{\textit{\v{S}}}_t$- and $\text{\textit{\c{C}}}_t$-equations, $w(L_t,\text{\textit{\c{C}}}_t)$ and $w(L_t,\text{\textit{\v{S}}}_t)$ are defined analogously. Sample period: 1973:01 to 2019:12.}
\end{figure}

% Fig 2: Loadings
The preceding discussion of $\tilde{\bm z}_t$ must be considered in light of the rescaled loadings in $\tilde{\bm \Lambda} = \bm \Lambda \bm U^{-1}$. $\tilde{\bm \Lambda}$ translates the law of motion captured in $\tilde{\bm z}_t$ to the coefficients in $\bm \beta_t$ by acting either as amplifier or attenuator. Figure \ref{fig:loadings} shows the posterior mean of the rescaled loadings $\bm \Lambda \bm U^{-1}$ and allows to assess which elements in $\tilde{\bm z}_t$ determine the time variation in the TVPs. We differentiate the loadings along two dimensions. Panel (a) shows the effect modifiers related to the VAR coefficients, while panel (b) depicts the block of $\tilde{\bm \Lambda}$ related to the covariances (stored in $\bm q_t$). 

Assessing the loadings in $\bm \Lambda_r$ for the $L_t$-equation reveals that a major part of the coefficients loads strongly positive on $r_{\text{NFCI},t}$. In the case of $r_{\text{RF},t}$ and $r_{\text{REC},t}$ the patterns are more mixed. For $r_{\text{NFCI},t}$ we find a maximum loading of around $0.25$ for $\text{\textit{\v{S}}}_{t-1}$, $L_{t-2}$, $\text{\textit{\c{C}}}_{t-2}$ and $L_{t-3}$, while for $r_{\text{RF},t}$ and $r_{\text{REC},t}$ some coefficients load moderately positive (e.g., the first own lag $L_{t-1}$) and others moderately negative (e.g., loadings related to lags of the curvature factor). The loadings $\bm \Lambda_S$ and $\bm \Lambda_\tau$ related to estimated latent factors exhibit only modest relevance for defining the law of motion in coefficients related to $L_t$. In particular, the amplifiers for $\bm{\tau}_t$ are shrunk heavily towards zero, implying that low frequency movements in the respective coefficients are either already captured by other measures, or irrelevant for coefficients in the $L_t$-equation. The same is true for the indicators in $\bm{S}_t$ and $\bm{\tau}_t$ for the case of the $\text{\textit{\v{S}}}_t$-equation. 

Turning to observed factors, the corresponding factor loadings are positive for lower-order lags while strongly negative for higher-order lags.
By contrast, we find mostly negative loadings for observed factors for first-lags in the $\text{\textit{\c{C}}}_t$-equation. Loadings related to higher-order lags are mixed, and no clear patterns are observable. Interestingly, the curvature factor is the only equation where we detect substantial loadings on the latent factors, with most loadings showing a positive sign.

While unobserved factors appear to be less important for coefficients in the conditional mean of the model, they play an important role for the covariances, as indicated in panel (b). The covariance between $\text{\textit{\v{S}}}_t$ and $\text{\textit{\c{C}}}_t$, $w(\text{\textit{\v{S}}}_t,\text{\textit{\c{C}}}_t)$, shows pronounced loadings on all other effect modifiers than NFCI, where we observe a modest negative loading. A mixed pattern emerges for the $L_t$ and $\text{\textit{\c{C}}}_t$ covariance, $w(L_t,\text{\textit{\c{C}}}_t)$, with some positive and negative measures. The contemporaneous relationship $w(L_t,\text{\textit{\v{S}}}_t)$ between $L_t$ and $\text{\textit{\v{S}}}_t$, marks a particularly interesting case, with modestly negative loadings on NFCI and RF, while REC and the unobserved factor loadings are close to zero.

Summarizing, we find that observed factors often load strongly on the coefficients of all equations. While the Markov switching indicator appears to be important particularly for the $L_t$- and $\text{\textit{\c{C}}}_t$-equations, loadings are muted for the coefficients of $\text{\textit{\v{S}}}_t$. Another interesting aspect is that conditional on observed effect modifiers, the gradually evolving coefficients captured by $\bm{\tau}_t$ are mostly irrelevant for the $L_t$- and $\text{\textit{\v{S}}}_t$-equations, different to strong positive loadings in the case of the $\text{\textit{\c{C}}}_t$-equation and the covariances. Additional results showing the actual regression coefficients are provided in Appendix \ref{app:betas}.

% Fig 3: Low frequency relationship 
\subsection{Low Frequency Relations between the Nelson-Siegel Factors}\label{sec: lowfrequency}
To assess what our framework implies on the relations between the factors that determine the yield curve, we compute the low-frequency relationship between the $L_t$, $\text{\textit{\v{S}}}_t$ and $\text{\textit{\c{C}}}_t$. We choose this long-run correlation measure for two reasons. First, it allows to illustrate movements in time-varying coefficients (i.e., transmission channels) and changes in the error variances in a single indicator over time. Second, this measure compresses information of all coefficients in a structurally meaningful way since it isolates long-run trends and correlations from short-run fluctuations \citep{sargent2011two, kliem2016low}. Additional results for the reduced form coefficients are provided in Appendix B.

To construct the measure, we transform the TVP-VAR($P$) in Eq. (\ref{eq: obs_states}) to its state-space TVP-VAR($1$) form. In what follows, the observation equation is given by $\bm y_t = \bm J \bm Y_t$, while the state equation is defined as $\bm Y_t = \bm B_t \bm Y_{t-1} + \bm \omega_t$ with $\bm \omega_t \sim \mathcal{N}(\bm 0, \bm \Omega_t)$. Here, $\bm J$ maps a $K$-dimensional vector $\bm Y_t = (\bm y_t', \dots, \bm y_{t-P+1}')'$ to $\bm y_t$, $\bm B_t$ collects the elements in $\bm \beta_t$ in the upper $M \times K$ block and defines identities otherwise. Similar, the $K \times K$-dimensional variance-covariance matrix $\bm \Omega_t$ collects elements in $\bm \Sigma_t$ in the upper-left $M \times M$ block and is zero otherwise. We follow \cite{sargent2011two} and first calculate the spectral density $\bm \Phi_t(0)$ of $\bm y_t$ at a zero frequency which coincides with the unconditional variance-covariance matrix of $\bm y_t$: 
\begin{equation*}
\bm \Phi_t(0) = \bm J \left(\bm I - \bm B_t \right)\bm \Omega_t (\bm I - \bm B'_t)^{-1}  \bm J', \quad \text{for } t = 1, \dots, T.
\end{equation*}
Next, we transform the covariances of $\bm \Phi_t$ into a correlation measure for each period $t$ and each variable combination $i,j~(i \neq j \text{ and } i,j = 1, \dots, M)$:
\begin{equation*}
\phi_{ij,t} = \frac{\Phi_{ij,t}(0)}{\Phi_{jj,t}(0)}.
\end{equation*}

\begin{figure}[!htbp]
\centering
\caption{Posterior median and the 68 percent credible set for pairwise long-run correlations.\label{fig:LFR}}
\begin{minipage}{\textwidth}
\centering
(a) $\phi_{L\text{\textit{\v{S}}},t}$
\end{minipage}
\begin{minipage}{\textwidth}
\centering
\hspace{2mm}%
\includegraphics[scale=0.42]{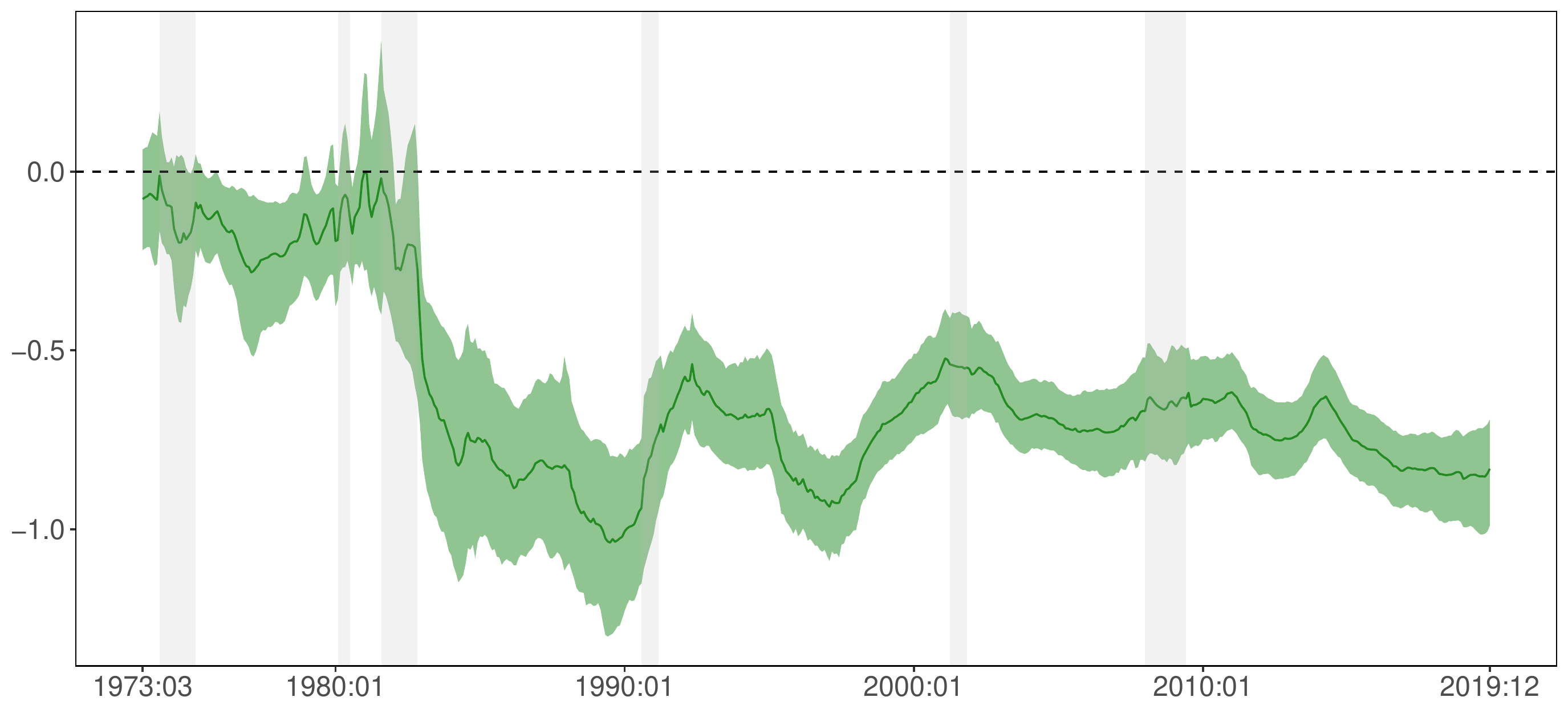}
\end{minipage}
\begin{minipage}{\textwidth}
\centering
\vspace{5pt}
(b) $\phi_{L\text{\textit{\c{C}}},t}$
\end{minipage}
\begin{minipage}{\textwidth}
\centering
\includegraphics[scale=0.42]{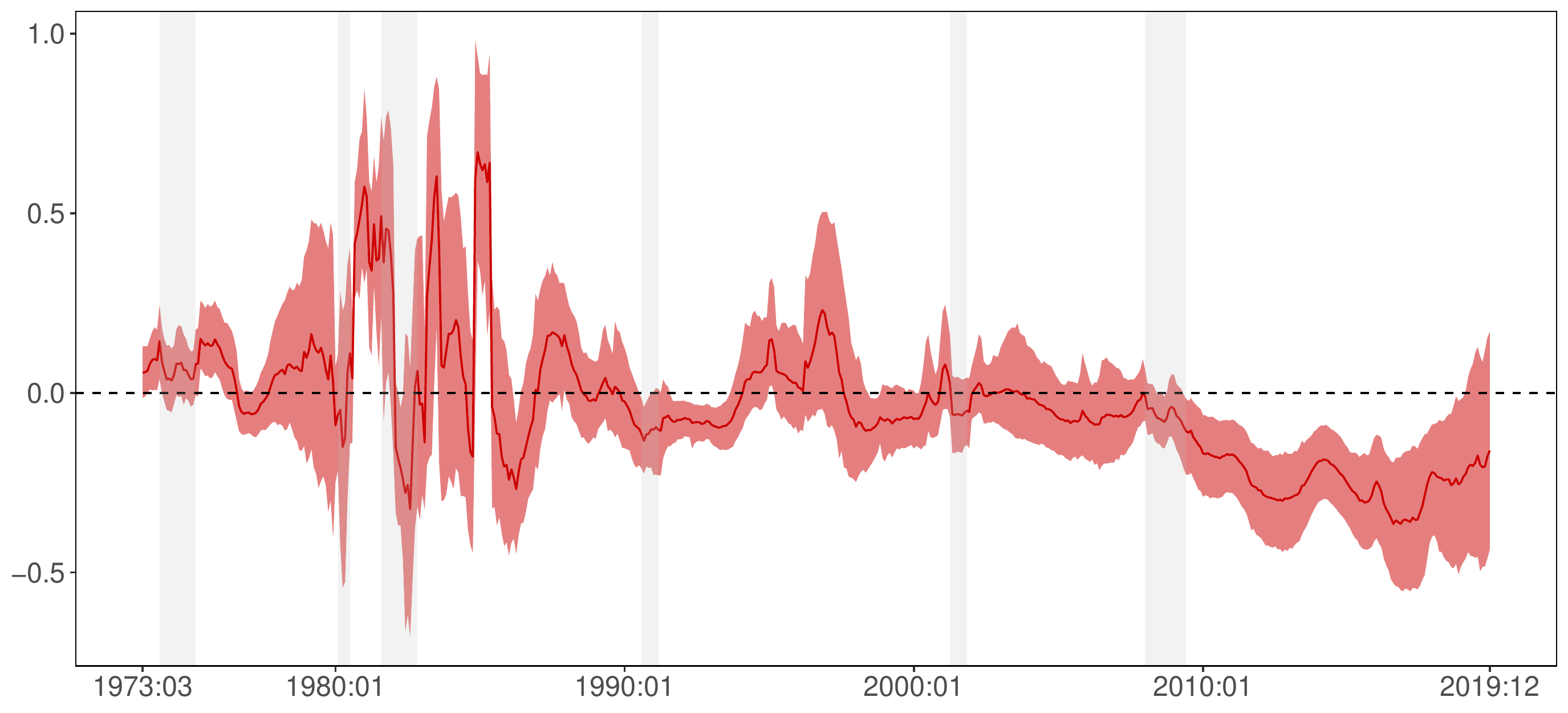}
\end{minipage}
\begin{minipage}{\textwidth}
\centering
\vspace{5pt}
(c) $\phi_{\text{\textit{\v{S}}}\text{\textit{\c{C}}},t}$
\end{minipage}
\begin{minipage}{\textwidth}
\centering
\includegraphics[scale=0.42]{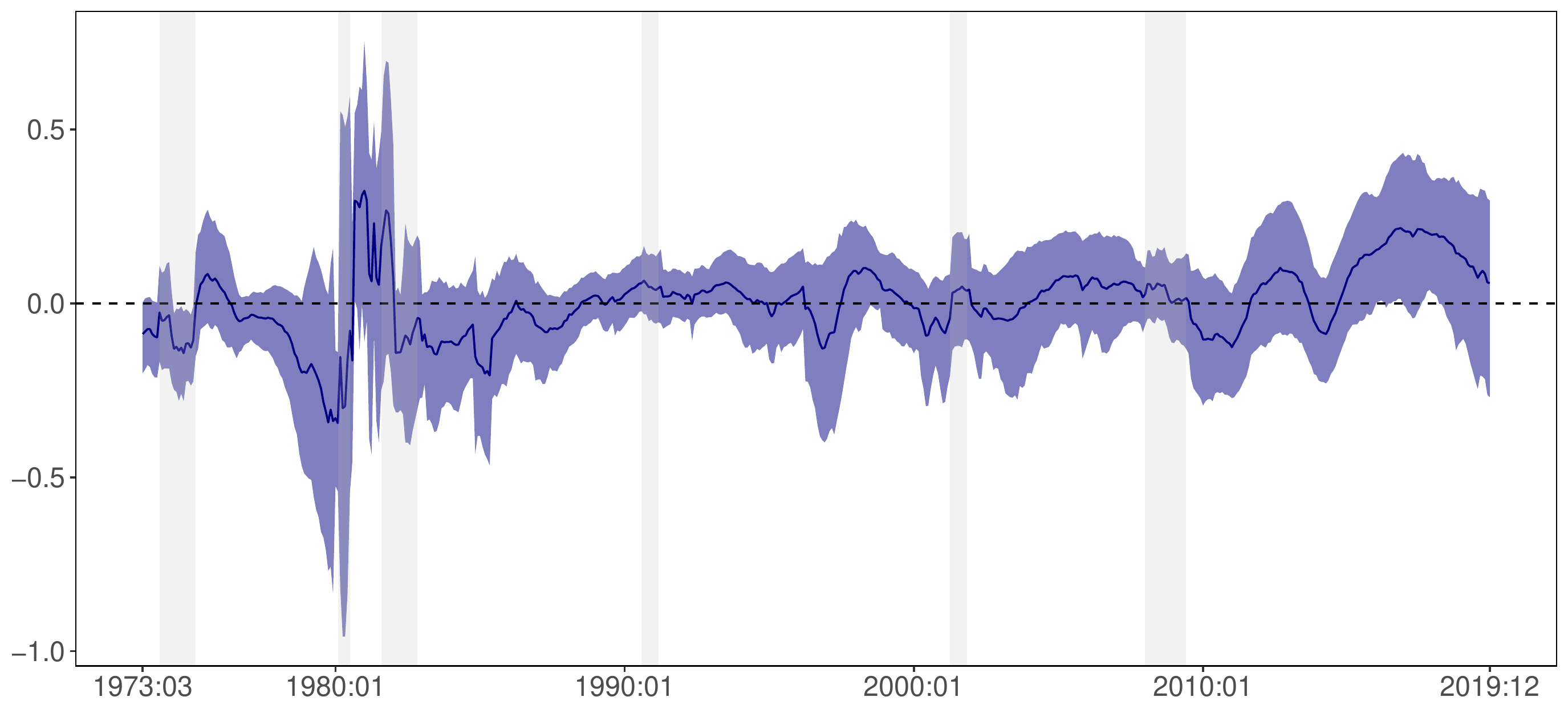}
\end{minipage}
\begin{minipage}{\textwidth}
\end{minipage}
\caption*{\footnotesize \noindent \textit{Notes}: Panel (a) $\phi_{L\text{\textit{\v{S}}},t}$, (b) $\phi_{L\text{\textit{\c{C}}},t}$ and (c) $\phi_{\text{\textit{\v{S}}}\text{\textit{\c{C}}},t}$, based on a TVP-NS-VAR with $\bm{z}_t=(\bm{r}^{\prime}_t,\bm{S}_t^\prime, \bm{\tau}_t^\prime)^\prime$, $\delta=R\tau/M=1$ and $P=3$. The colored solid lines denote the posterior medians, the black dashed line the zero line, and the colored shaded band the 68 percent posterior coverage interval, while the gray vertical bars represent recessions dated by the NBER Business Cycle Dating Committee. $\phi_{ij,t}$ is the long-run correlation for the variables $i$ and $j$ at time $t$, for $i, j \in \{L, \text{\textit{\v{S}}}, \text{\textit{\c{C}}}\}$. Results are based on 15,000 MCMC draws. Sample period: 1973:01 to 2019:12. Vertical axis: correlation measurements. Front axis: months.  }
\end{figure}

The measure $\phi_{ij,t}$ describes the long-run relations between variable $i$ and $j$ at each point in time, and is displayed in Figure \ref{fig:LFR} for the model specification TVP-NS-VAR with $\bm{z}_t=(\bm{r}_t^\prime, \bm{S}_t^\prime, \bm{\tau}_t^\prime)^\prime$, $R_{\tau j}=1$ and $P=3$. Note that the variables enter our model in differences. Hence, Figure \ref{fig:LFR} depicts the low frequency relations of \textit{changes} in the level, slope and curvature of the yield curve.

We observe several interesting periods characterized by structural breaks. First, the relationship between the level and slope of the yield curve was close to zero until the Volcker disinflation of the 1980s. After this period, we can identify an abrupt decrease to significantly negative values. The long-run correlation stays negative until the end of the sample, with minor low-frequency movements. 

Second, for most of the sample the long-run coefficient between level and curvature of the yield curve is insignificant. Substantial structural breaks are detectable in the early 1980s. Again, this coincides with a shift in the US monetary policy regime. Between the two recessions in the early 1980s and in the recovery period afterwards, we note a strong positive relationship between the level and curvature of the yield curve. Afterwards, during the Great Moderation, there are mostly insignificant values. This ends during the Great Recession, where the relationship is estimated to be significantly negative. This finding may be linked to short-term interest rates approaching the zero lower bound rapidly. The trend reverses late in the sample, with the Federal Reserve conducting several subsequent rate hikes starting at the end of 2015. Turning to the relationship between the slope and curvature of the yield curve, we again find an insignificant relationship for most of the sample. Large breaks are observable in the period between the two 1980s recessions, but different to the relationship between the level and curvature of the yield curve, this is not visible in the subsequent recovery period. Interestingly, we estimate a significantly positive relationship in the period of the aforementioned rate hikes by the Federal Reserve starting in 2015.

\section{Conclusions}\label{sec:conclusion}
This paper proposes methods for automatically selecting adequate state equations in TVP-VAR models in a data-driven fashion. The TVPs are assumed to depend on a set of observed and unobserved covariates, also known as effect modifiers. As unobserved covariates, we consider a set of low dimensional latent factors that follow a random walk, alongside Markov switching indicators that allow for abrupt structural breaks. Our model nests several alternatives commonly used in the literature on modeling macroeconomic and financial time series. To choose between state equations, we use a hierarchical Bayesian global-local shrinkage prior on the most flexible specification.

We apply our econometric framework to US yield curve data. Carrying out a thorough predictive exercise, we show that our techniques produce favorable point and density forecasts vis-\`{a}-vis a set of established benchmark models (which are nested variants of our proposed modeling approach). The performance is specific to the information set used in the underlying TVP-VAR and appears to be more pronounced for density forecasts. This exercise illustrates that our approach produces very competitive forecasts without increasing the risk of overfitting, while providing a framework to trace the sources of time-variation -- a key advantage compared to conventional TVP-VARs. This predictive exercise is complemented by a full-sample analysis of structural breaks in the relationship between the level, slope and curvature of the US yield curve. We detect several interesting patterns in abrupt and gradual time-variation patterns in long-run cross-variable relations. These changes appear to be specific to the monetary regime and the state of the business cycle.\\

\noindent \textbf{Acknowledgments}: The authors gratefully acknowledge financial support by the Jubil\"aums-fonds of the Oesterreichische Nationalbank (OeNB, project 18127) and by the Austrian Science Fund (FWF, project ZK 35).

\small{\setstretch{2}
\addcontentsline{toc}{section}{References}
\bibliographystyle{plainnat}
\bibliography{tvp}}\normalsize

\newpage
\begin{appendices}
\begin{center}
\LARGE\textbf{Appendices}
\end{center}

\setcounter{equation}{0}
\renewcommand\theequation{A.\arabic{equation}}
\section{Technical appendix}\label{app:technical}
\subsection{Sampling the state innovation variances}
For sampling the state innovation variances based on Eq. (\ref{eq: obs_states}), we let ${\eta}_{ji,t}$ denote the shock to the $i^{\text{th}}$ coefficient in $\tilde{\bm{\gamma}}_t$ with respect to the $j^{\text{th}}$ equation. The posterior of the state innovation variances is a generalized inverse Gaussian (GIG) distribution:\footnote{The generalized inverse Gaussian distribution is specified such that its density function is proportional to $f(x)=x^{\lambda-1}\exp(-(\chi/x + \psi x)/2)$ for a random variable $x\sim\mathcal{GIG}(\lambda,\chi,\psi)$.}
\begin{equation*}
\omega_{ji}|\varpi_{ji},\vartheta_j\sim\mathcal{GIG}\left(\frac{1-T}{2},\sum_{t=1}^T \eta_{ji,t}^2,\varpi_{ji}\vartheta_j\right).
\end{equation*}

\subsection{Posterior for the horseshoe prior}
Our specification of the horseshoe prior on $\bm{\Lambda}_j$, $\bm{\gamma}_j$ and the square root of $\bm{\omega}_j$ in Section \ref{prior} for a generic parameter $b_i$ for $i=1,\hdots,K$ is:
\begin{equation*}
b_i|c_i,d \sim \mathcal{N}(0,c_i^2d^2),\quad c_i\sim\mathcal{C}^{+}(0,1),\quad d\sim\mathcal{C}^{+}(0,1).
\end{equation*}
We rely on this prior in its auxiliary representation as in \citet{makalic2015simple} for efficient sampling of the local ($c_i$) and global ($d$) shrinkage parameters:
\begin{equation*}
c_i^2|e_i\sim\mathcal{G}^{-1}(1/2,1/e_i),\quad d^2|f\sim\mathcal{G}^{-1}(1/2,1/f), \quad e_i\sim\mathcal{G}^{-1}(1/2,1), \quad f\sim\mathcal{G}^{-1}(1/2,1).
\end{equation*}
Here, $\mathcal{G}^{-1}$ denotes the inverse Gamma distribution. This setup yields the following conditional posterior distributions:
\begin{align*}
c_i^2|b_i,d,e_i&\sim\mathcal{G}^{-1}\left(1,\frac{1}{e_i}+\frac{b_i^2}{2d^2}\right), \quad d^2|b_i,c_i,f\sim\mathcal{G}^{-1}\left(\frac{K+1}{2},\frac{1}{f}+\sum_{i=1}^K \frac{b_i^2}{2c_i^2}\right),\\
e_i|c_i&\sim\mathcal{G}^{-1}\left(1,1+c_i^{-2}\right),\quad f|d\sim\mathcal{G}^{-1}\left(1,1+d^{-2}\right).
\end{align*}

\setcounter{equation}{0}
\renewcommand\theequation{B.\arabic{equation}}
\setcounter{figure}{0}
\renewcommand\thefigure{B.\arabic{figure}}    
\section{Further empirical results}\label{app:betas}
This Appendix contains additional results for the reduced form coefficients. While Sub-section \ref{sec:insample} provides posterior estimates of the lower dimensional effect modifiers, Figs. \ref{fig:betasown}--\ref{fig:contemp} display the coefficients obtained by multiplying the factor loadings with the observed/latent factors based on the relationship established in Eq. (\ref{eq: obs_states}).

\setcounter{figure}{0}
\renewcommand\thefigure{B.\arabic{figure}}    

\begin{figure}[!htbp]
\centering
\caption{Posterior median of the coefficients associated with the own lags of $L_t$, $\text{\textit{\v{S}}}_t$ and $\text{\textit{\c{C}}}_t$. \label{fig:betasown}}
\begin{minipage}{\textwidth}
\centering
(a) \textit{Coefficients of first own lag}
\end{minipage}
\begin{minipage}{\textwidth}
\centering
\includegraphics[scale=0.42]{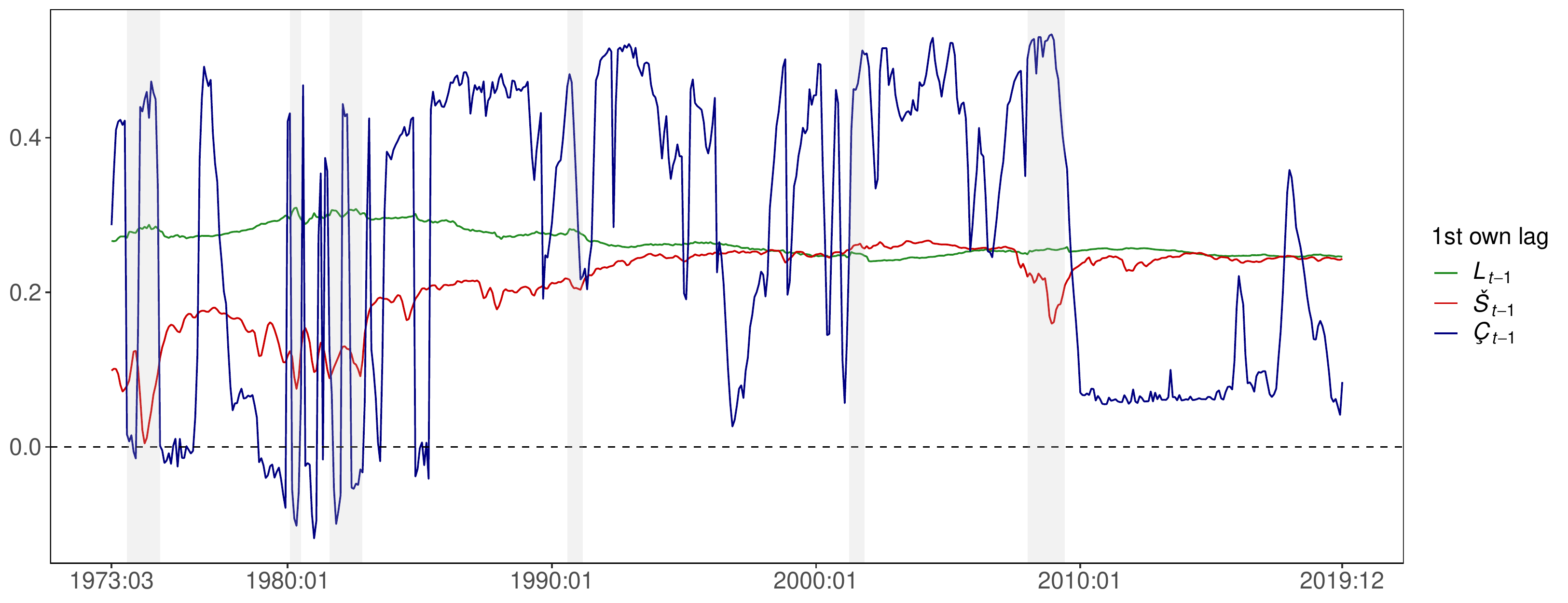}
\end{minipage}
\begin{minipage}{\textwidth}
\centering
\vspace{5pt}
(b) \textit{Coefficients of second own lag}
\end{minipage}
\begin{minipage}{\textwidth}
\centering
\includegraphics[scale=0.427]{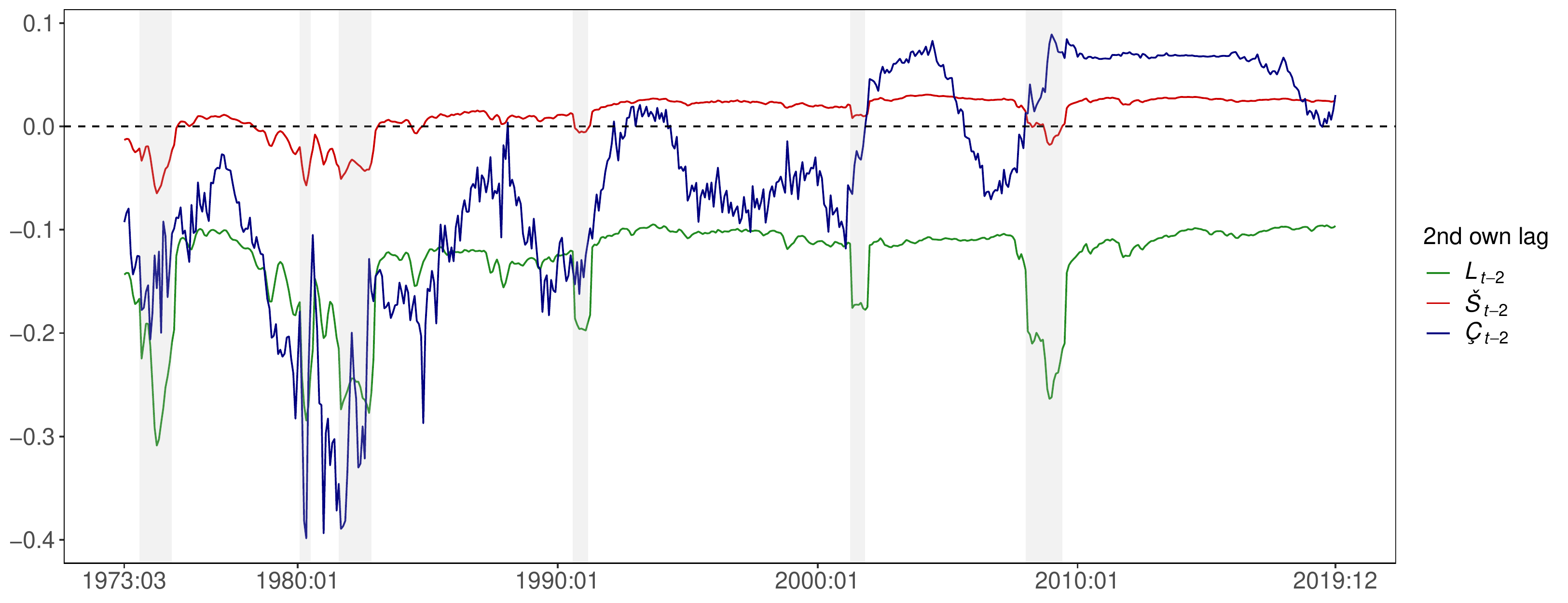}
\end{minipage}
\begin{minipage}{\textwidth}
\centering
\vspace{5pt}
(c) \textit{Coefficients of third own lag}
\end{minipage}
\begin{minipage}{\textwidth}
\centering
\advance\leftskip-3pt
\includegraphics[scale=0.43]{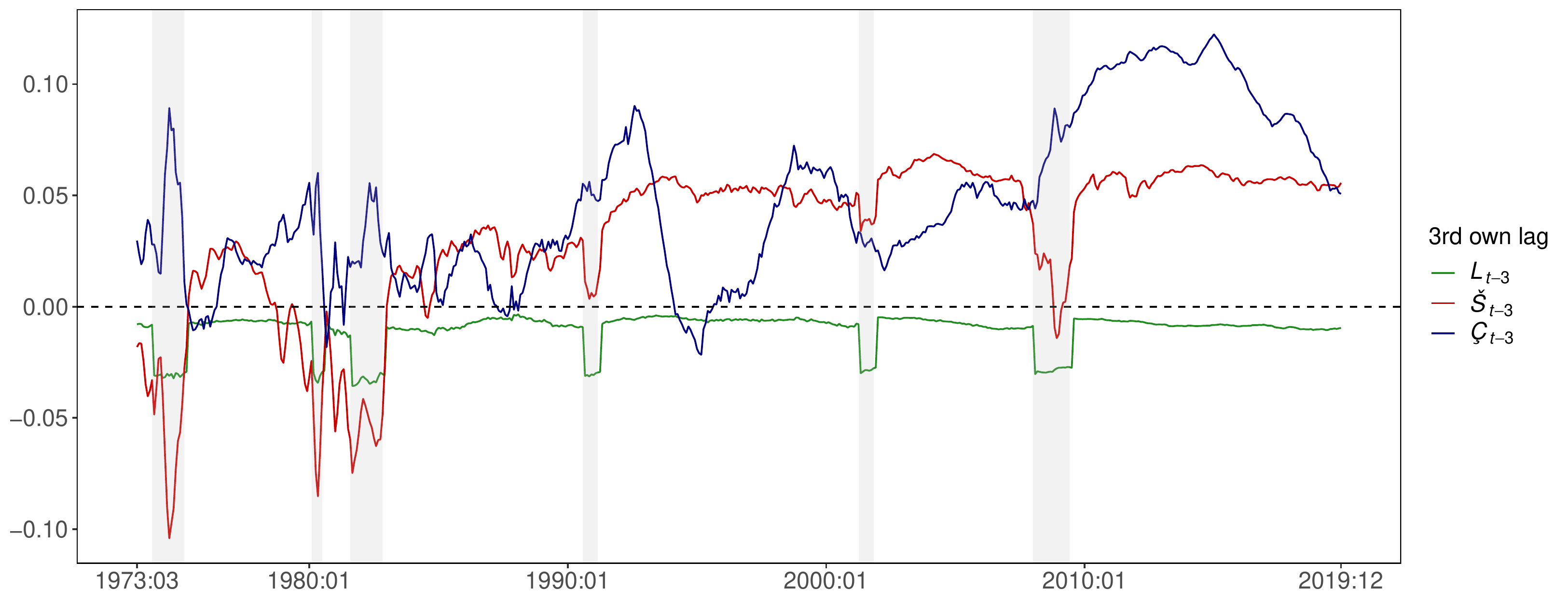}
\end{minipage}
\begin{minipage}{\textwidth}
\end{minipage}
\caption*{\footnotesize \noindent \textit{Notes}: Panels (a), (b) and (c) show the dynamic evolution of the coefficients related to the variables' own lags $p\in\{1,2,3\}$ of the respective equation for $L_t$, $\text{\textit{\v{S}}}_t$ and $\text{\textit{\c{C}}}_t$. Results are based on the TVP-NS-VAR model variant with $\delta= R_\tau/M=1$ and using 15,000 MCMC draws. The black dashed line denotes the zero line, while the gray shaded vertical bars represent recessions dated by the NBER Business Cycle Dating Committee. Sample period 1973:01 to 2019:12. Vertical axis: posterior median estimate. Front axis: months.}
\end{figure}

\begin{figure}[!htbp]
\centering
\caption{Posterior median of the coefficients associated with cross-variable lags of $L_t$, $\text{\textit{\v{S}}}_t$ and $\text{\textit{\c{C}}}_t$. \label{fig:betasother}}
\begin{minipage}{\textwidth}
\centering
(a) \textit{Coefficients of first other lags}
\end{minipage}
\begin{minipage}{\textwidth}
\centering
\includegraphics[scale=0.42]{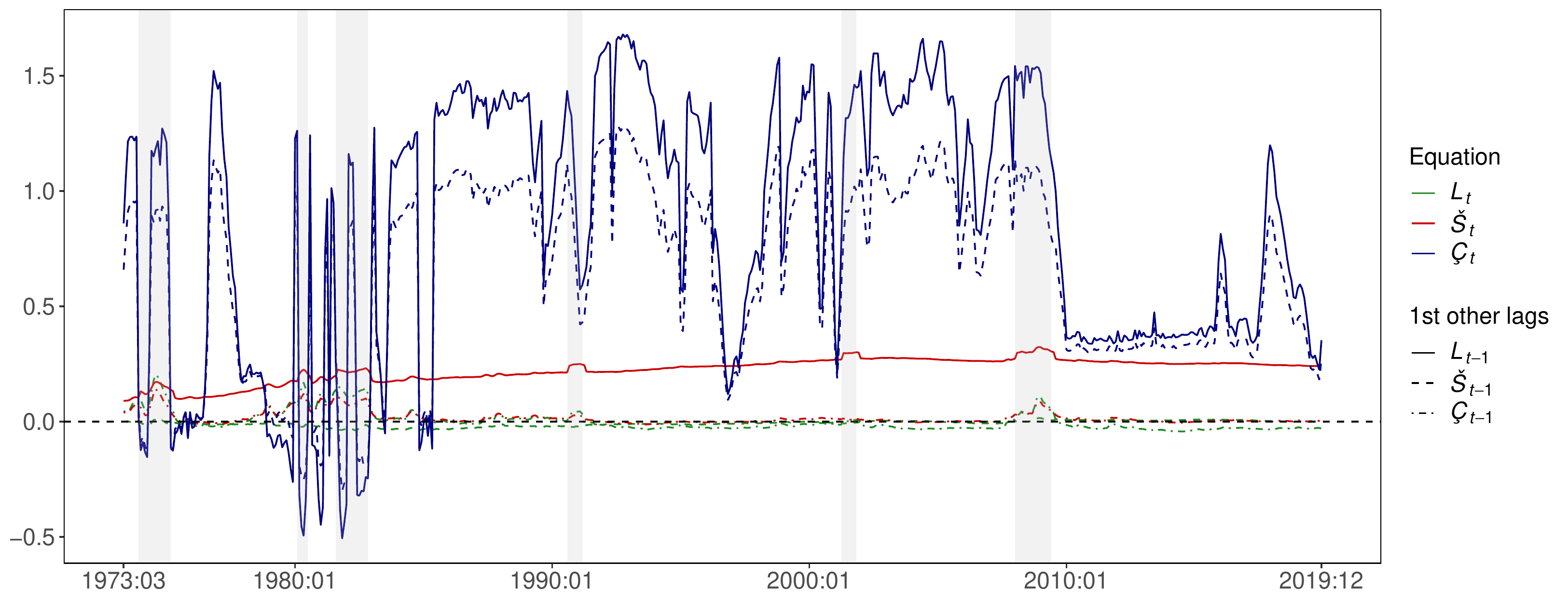}
\end{minipage}
\begin{minipage}{\textwidth}
\centering
\vspace{5pt}
(b) \textit{Coefficients of second other lags}
\end{minipage}
\begin{minipage}{\textwidth}
\centering
\hspace{0.05pt}
\includegraphics[scale=0.423]{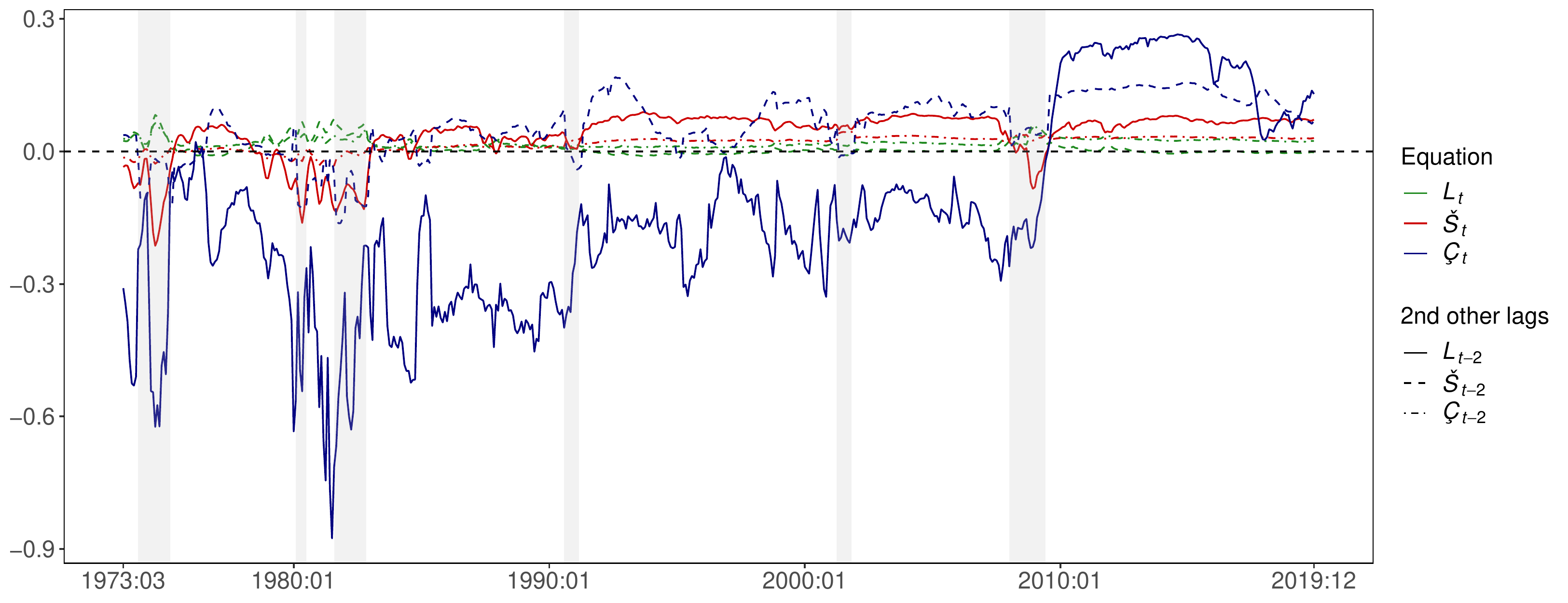}
\end{minipage}
\begin{minipage}{\textwidth}
\centering
\vspace{5pt}
(c) \textit{Coefficients of third other lags}
\end{minipage}
\begin{minipage}{\textwidth}
\centering
\hspace{0.08pt}
%\advance\leftskip0.08pt
\includegraphics[scale=0.42]{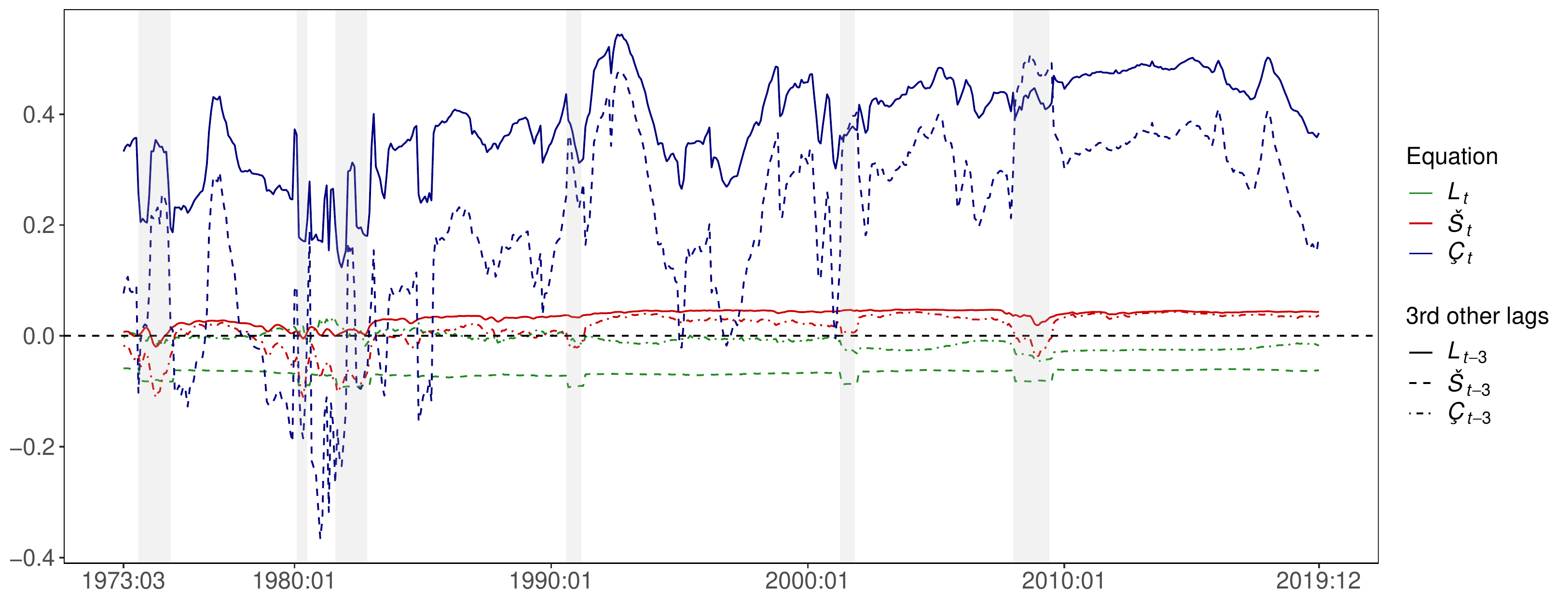}
\end{minipage}
\begin{minipage}{\textwidth}
\end{minipage}
\caption*{\footnotesize \noindent \textit{Notes}: Panels (a), (b) and (c) show the dynamic evolution of the coefficients related to cross-variable lags $p\in\{1,2,3\}$ of the respective equation for $L_t$, $\text{\textit{\v{S}}}_t$ and $\text{\textit{\c{C}}}_t$. Results are based on the TVP-NS-VAR model variant with $\delta= R_\tau/M=1$ and using 15,000 MCMC draws. The black dashed line denotes the zero line, while the gray shaded vertical bars represent recessions dated by the NBER Business Cycle Dating Committee. Sample period 1973:01 to 2019:12. Vertical axis: posterior median estimate. Front axis: months.}
\end{figure}

\newpage
\clearpage

\begin{figure}[!htbp]
\centering
\caption{Posterior median of the contemporaneous relationships between $L_t$, $\text{\textit{\v{S}}}_t$ and $\text{\textit{\c{C}}}_t$. \label{fig:contemp}}
\begin{minipage}{\textwidth}
\centering
\includegraphics[scale=0.42]{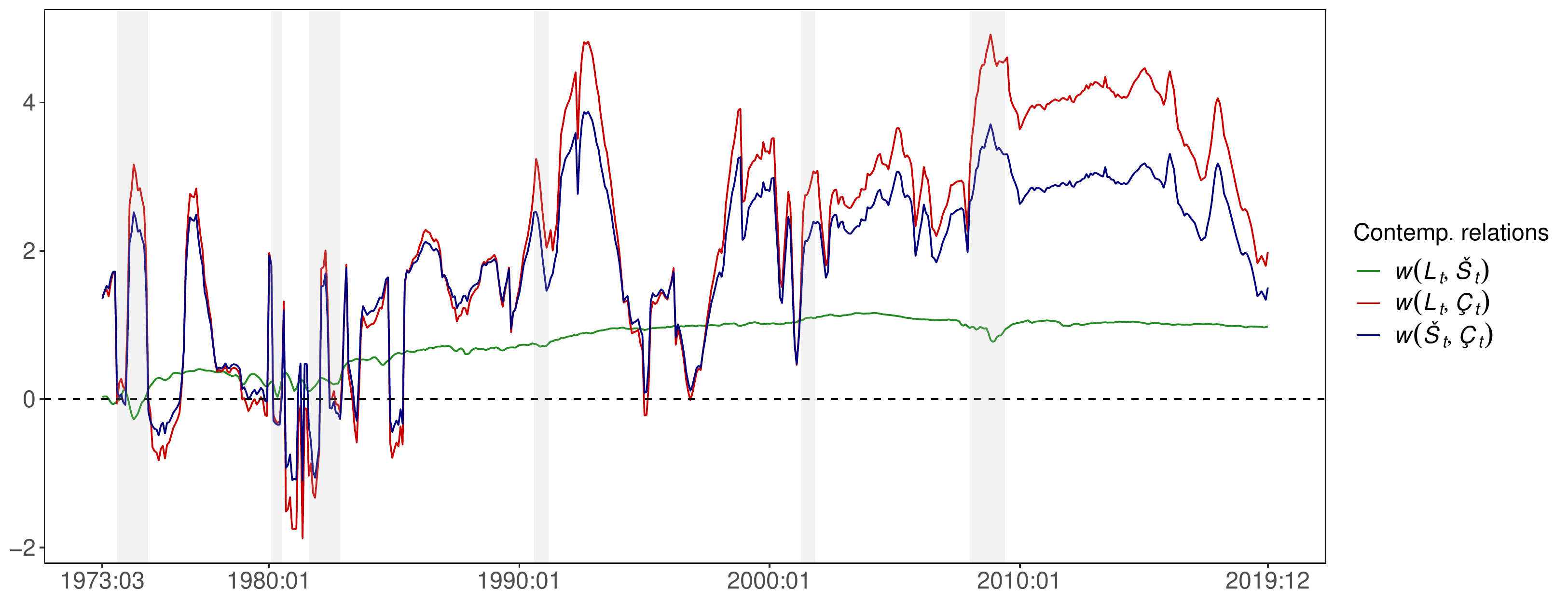}
\end{minipage}
\begin{minipage}{\textwidth}
\end{minipage}
\caption*{\footnotesize \noindent \textit{Notes}: $w(\text{\textit{\v{S}}}_t, \text{\textit{\c{C}}}_t)$ denotes the contemporaneous relation between the $\text{\textit{\v{S}}}_t$- and $\text{\textit{\c{C}}}_t$-equations, $w(L_t,\text{\textit{\c{C}}}_t)$ and $w(L_t,\text{\textit{\v{S}}}_t)$ are defined analogously. Results are based on the TVP-NS-VAR model variant with $\delta=R_\tau/M=1$ and using 15,000 MCMC draws. The black dashed line denotes the zero line, while the gray shaded vertical bars represent recessions dated by the NBER Business Cycle Dating Committee. Sample period 1973:01 to 2019:12. Vertical axis: posterior median estimate. Front axis: months.}
\end{figure}

\end{appendices}

\end{document}